\documentclass[preprint2]{aastex}

\slugcomment{Not to appear in Nonlearned J., 45.}

\shorttitle{GJ436b IRAC ch1 and ch2}
\shortauthors{Morello et al.}

\begin{document}

%% LaTeX will automatically break titles if they run longer than
%% one line. However, you may use \\ to force a line break if
%% you desire.

\title{Revisiting Spitzer transit observations with Independent Component Analysis: new results for the GJ436 system}

%% Use \author, \affil, and the \and command to format
%% author and affiliation information.
%% Note that \email has replaced the old \authoremail command
%% from AASTeX v4.0. You can use \email to mark an email address
%% anywhere in the paper, not just in the front matter.
%% As in the title, use \\ to force line breaks.

\author{G. Morello, I. P. Waldmann, G. Tinetti, I. D. Howarth}
\affil{Department of Physics \& Astronomy, University College London, Gower Street, WC1E6BT, UK}
\email{giuseppe.morello.11@ucl.ac.uk}

\author{G. Micela}
\affil{INAF -  Osservatorio Astronomico di Palermo, Piazza del Parlamento 1, 90134, Italy}

\and

\author{F. Allard}
\affil{Centre de Recherche Astrophysique  de Lyon - \'Ecole Normale Sup\'erieure de Lyon, 46 All\'ee d'Italie, 69364 Lyon Cedex 07}

%% Notice that each of these authors has alternate affiliations, which
%% are identified by the \altaffilmark after each name.  Specify alternate
%% affiliation information with \altaffiltext, with one command per each
%% affiliation.

%% Mark off your abstract in the ``abstract'' environment. In the manuscript
%% style, abstract will output a Received/Accepted line after the
%% title and affiliation information. No date will appear since the author
%% does not have this information. The dates will be filled in by the
%% editorial office after submission.

\begin{abstract}
We analyzed four Spitzer/IRAC observations at 3.6 and 4.5 $\mu$m of the primary transit of the exoplanet GJ436b, by using blind source separation techniques. These observations are important to investigate the atmospheric composition of the planet GJ436b. Previous analyses claimed strong inter-epoch variations of the transit parameters due to stellar variability, casting doubts on the possibility to extract conclusively an atmospheric signal; those analyses also reported discrepant results, hence the necessity of this reanalysis. The method we used has been proposed in \cite{mor14} to analyze 3.6 $\mu$m transit light-curves of the hot Jupiter HD189733b; it performes an Independent Component Analysis (ICA) on a set of pixel-light-curves, i.e. time series read by individual pixels, from the same photometric observation. Our method only assumes the independence of instrumental and astrophysical signals, and therefore guarantees a higher degree of objectivity compared to parametric detrending techniques published in the literature. The datasets we analyzed in this paper represent a more challenging test compared to the previous ones. \\
Contrary to previous results reported in the literature, our results (1) do not support any detectable inter-epoch variations of orbital and stellar parameters, (2) are photometrically stable at the level $\sim$10$^{-4}$ in the IR, and (3) the transit depth measurements at the two wavelengths are consistent within 1$\sigma$. We also (4) detect a possible transit duration variation (TDV) of $\sim$80 s (2 $\sigma$ significance level), that has not been pointed out in the literature, and (5) confirm no transit timing variations (TTVs) $\gtrsim$30 s. 
\end{abstract}

%% Keywords should appear after the \end{abstract} command. The uncommented
%% example has been keyed in ApJ style. See the instructions to authors
%% for the journal to which you are submitting your paper to determine
%% what keyword punctuation is appropriate.


%% From the front matter, we move on to the body of the paper.
%% In the first two sections, notice the use of the natbib \citep
%% and \citet commands to identify citations.  The citations are
%% tied to the reference list via symbolic KEYs. The KEY corresponds
%% to the KEY in the \bibitem in the reference list below. We have
%% chosen the first three characters of the first author's name plus
%% the last two numeral of the year of publication as our KEY for
%% each reference.

%% Authors who wish to have the most important objects in their paper
%% linked in the electronic edition to a data center may do so by tagging
%% their objects with \objectname{} or \object{}.  Each macro takes the
%% object name as its required argument. The optional, square-bracket 
%% argument should be used in cases where the data center identification
%% differs from what is to be printed in the paper.  The text appearing 
%% in curly braces is what will appear in print in the published paper. 
%% If the object name is recognized by the data centers, it will be linked
%% in the electronic edition to the object data available at the data centers  
%%
%% Note that for sources with brackets in their names, e.g. [WEG2004] 14h-090,
%% the brackets must be escaped with backslashes when used in the first
%% square-bracket argument, for instance, \object[\[WEG2004\] 14h-090]{90}).
%%  Otherwise, LaTeX will issue an error. 

\section{Introduction}
Transit spectroscopy and differential photometry are largely used to investigate the composition and structure of exoplanetary atmospheres. The large majority of transiting exoplanets are ``hot Jupiters'', i.e. planets with size similar to Jupiter orbiting very closely to their host star (semimajor axis $\sim 0.01 - 0.5$AU). Their typical surface temperatures are $\gtrsim 1000$K.

GJ436b is a Neptune-sized planet orbiting around an M dwarf  with radius $\sim$0.46$R_{\odot}$ at a distance $\sim$0.03 AU. This planet is interesting for several reasons. It is one of the smallest  (radius $\sim$4.3$R_{\oplus}$) and coolest ($\sim 700$K) exoplanet for which optical-to-IR spectra have been measured \citep{gil07, dem07, alo08, cou08, cac09, dem09, pont09, bal10, ste10, bea11, knu11, knu14}. The primary transit depth is $\sim$0.7$\%$. Another peculiarity of GJ436b is its high orbital eccentricity ($e \sim$0.16), inferred from radial velocity measurements \citep{man07} and from secondary eclipse phasing \citep{dem09}. Both the physical and dynamical properties of GJ436b are debated in the literature.

\cite{man07} and \cite{dem07} investigated the origin of the high orbital eccentricity of GJ436b, concluding that the circularization timescale ($\sim$10$^8$ yr) is significantly smaller than the age of the system ($\gtrsim$6$\times$10$^9$). \cite{man07} also found a long trend in radial velocity measurements; they suggested the presence of an external perturber on a wider orbit to explain both the high eccentricity of GJ436b and the long trend in radial velocity measurements. \cite{rib08} hypothesized a Super-Earth on a close orbit to explain those evidences, later retracted. Transit timing variations (TTVs) reported by \cite{alo08, cac09} do not support any evidence of external perturbers. \cite{ste12} claimed the possible detection of two nearby sub-Earth-sized exoplanets transiting in GJ436 system; according to the authors, the dynamic of the proposed system is consistent with the current non-TTV-detections.

Based on multiwavelength infrared eclipse measurements, \cite{ste10} proposed a high CO-to-CH$_4$ ratio compared to thermochemical equilibrium models for hydrogen-dominated atmospheres. Their atmospheric model includes disequilibrium processes, such as vertical mixing and polymerization of methan to explain the observed deficiency of CH$_4$. \cite{bea11} suggested strong CH$_4$ absorption at 3.6, 4.5, and 8.0 $\mu$m Spitzer/IRAC passbands from primary transit observations, and their reanalysis of secondary eclipse data is consistent with this detection. \cite{knu11} measured significant time variations of the transit depths at the same wavelengths, which strongly affect the inferred transmission spectrum. They attributed such variations to the stellar activity and found that different results are obtainable depending on the observations considered. By rejecting those observations that they believe to be most strongly affected by stellar activity, their final results support CO as the dominant carbon molecule, with very little, if any, CH$_4$. More recent Hubble/WFC3 observations in the 1.2$-$1.6 $\mu$m wavelength interval, analyzed by \cite{knu14}, indicate a featureless transmission spectrum, which is consistent with relatively hydrogen-poor atmosphere with a high cloud or haze layer.

In this paper we reanalyze four transit light-curves obtained with Spitzer/IRAC at 3.6 and 4.5 $\mu$m passbands (channels 1 and 2 of IRAC). We adopt a non-parametric data detrending technique, based on Independent Component Analysis (ICA) applied to single pixel-light-curves, to ensure a higher degree of objectivity. This method has proven to give robust results, when applied to the transits of the hot-Jupiter HD189733b observed with IRAC at 3.6 $\mu$m \citep{mor14}. We further test here the performance of this detrending technique with the more challenging datasets of the Neptune-sized planet GJ436b, for which the transit depth is comparable with the amplitude of the instrumental pixel-phase signal, and the transit duration is very similar to the period of that signal. Additionally, we discuss the stellar and orbital stability of the GJ436 system, the repeatability of transit measurements, potentially affected by stellar, planet, and instrument variability, and the atmospheric contribution. We discuss the reliability of our results in light of other observations reported in the literature, in particular \cite{bea11, knu11, knu14}.

\section{Data Analysis}

\subsection{Observations}

We analyze four photometric observations of GJ436b, which are part of the Spitzer program ID 50051. They include two 3.6 and two 4.5 $\mu$m primary transits as detailed in Tab. \ref{tab1}.
\begin{table*}[t]
\begin{center}
\caption{Spitzer observations of primary transits of GJ436b. \label{tab1}}
\begin{tabular}{ccccc}
\tableline\tableline
Obs. Number & Detector & Wavelength ($\mu$m) & UT Date & Orbit Number\\
\tableline
1a & IRAC, ch1 & 3.6 & 2009 Jan 9 & 234\\
1b & IRAC, ch1 & 3.6 & 2009 Jan 28 & 241\\
2a & IRAC, ch2 & 4.5 & 2009 Jan 17 & 237\\
2b & IRAC, ch2 & 4.5 & 2009 Jan 31 & 242\\
\tableline
\end{tabular}
%% Any table notes must follow the \end{tabular} command.
\end{center}
\end{table*}
Each observation consists of 1829 exposures using IRAC's sub-array mode, taken over 4.3 hr: 0.8 hr on the primary transit of the planet, the remaining 3.5 hr before and after transit. The interval between consecutive exposures is 8.4 s. Each exposure includes 64 consecutive frames integrated over 0.1 s. We replaced the single frames of each exposure with their averages to reduce the random scatter, and the computational time\footnote{Computational time is dominated by the time for transit fitting, which strongly depends on the algorithms (and settings) used. In our case, transit fitting time was of several hours, and it scales as $\mathcal{O}(dN)$, being $d$ the number of free transit parameters, and $N$ the data points.}. During an observation, the centroid of the star GJ436 is stable to within one pixel.

\subsection{Detrending method, light-curve fitting and error bars}
\label{ssec:errors}

Here we outline the main steps of the analysis, i.e. data detrending, light-curve fitting and estimating parameter error bars. Further details are reported in \cite{mor14}.

To detrend the transit signals from single observations, we performed an ICA decomposition over selected pixel-light-curves, i.e. time series from individual pixels. We considered 5$\times$5 arrays of pixels with the stellar centroids at their centers. In this way, we obtain a set of maximally independent components: one of them is the transit signal, others may be instrumental systematics and/or astrophysical signals. Observed light-curves are linear combinations of these independent components, the coefficients of the linear combinations can be calculated by fitting the out-of-transit parts. To estimate the transit signal in a robust way, the fit is performed on the out-of-transit of the relevant integral light-curve, i.e. the sum of the pixel-light-curves from the array used, including all the non-transit components plus a constant term. The detrended transit signal is obtained by subtracting all the non-transit components, properly scaled by their fitting coefficients, from the integral light-curve. It is renormalized by the mean value on the out-of-transit, so that the out-of-transit level is unity.

After the extractions of the detrended and normalized transit time series, we modelled them by using the \cite{ma02} analytical formulae. We originally assumed the orbital period, $P$, and the epoch of the first transit, $E_{tr}$, reported by \cite{cac09}; the eccentricity, $e$, and the argument of periastron, $\omega$, reported by \cite{man07}; they are consistent with those reported by previous papers \citep{but04, gil07, dem09}, and more accurate. We tested two different sets of quadratic limb darkening coefficients for the star \citep{how11b}, $\gamma_1$ and $\gamma_2$, derived by an Atlas \citep{kur70, how11} and a Phoenix \citep{all95, all01} models  (see Sec. \ref{ssec:ldc}). With these settings, we estimated the planet-to-star radii ratio, $p = \frac{r_p}{R_s}$, the orbital semimajor axis in units of stellar radii, $a_0 = \frac{a}{R_s}$, and inclination, $i$. First estimates were obtained through a Nelder-Mead optimization algorithm \citep{lag98}; they were used as optimal starting points for an Adaptive Metropolis algorithm with delayed rejection \citep{haa06}, generating chains of 20,000 values. Updated best estimates and (partial) error bars of the parameters, $\sigma_{par,0}$, are the means and standard deviations of the relevant (gaussian distributed) sampled chains, respectively. The final parameter error bars are:
\begin{equation}
\label{eqn:sigmapar}
\sigma_{par} = \sigma_{par,0} \sqrt{ \frac{ \sigma_{0}^2 + \sigma_{ICA}^2 }{ \sigma_{0}^2}}
\label{eqn:sigma_par}
\end{equation}
$\sigma_{0}^2$ is the sampled likelihood variance, approximately equal to the variance of the residuals for the best transit model; $\sigma_{ICA}^2$ is a term estimating the uncertainty associated to the ICA extraction (see App. \ref{sec:app0} for further details). 

For completeness, and for comparison with the literature, we also calculated the transit depth, $p^2$, the impact parameter, $b$, and the transit duration, $T$, where (see \cite{ford08}):
\begin{equation}
b = a_0 \cos{i} \frac{1-e^2}{1+e \sin{ \omega}}
\end{equation}
\begin{equation}
T = \frac{P \sqrt{1-b^2}}{ \pi a_0} \frac{ \sqrt{1-e^2}}{1+e \sin{ \omega}}
\label{eqn:T}
\end{equation}

For a more thorough analysis, we performed other fits with different choices of the free parameters, introducing a phase-shift parameter to consider possible timing error/variations, and simultaneous fits on more than one multiple light-curves with some common free parameters.

\subsection{Application to observations}
\label{ssec:application}

\begin{figure}[!h]
%%\epsscale{0.80}
\plotone{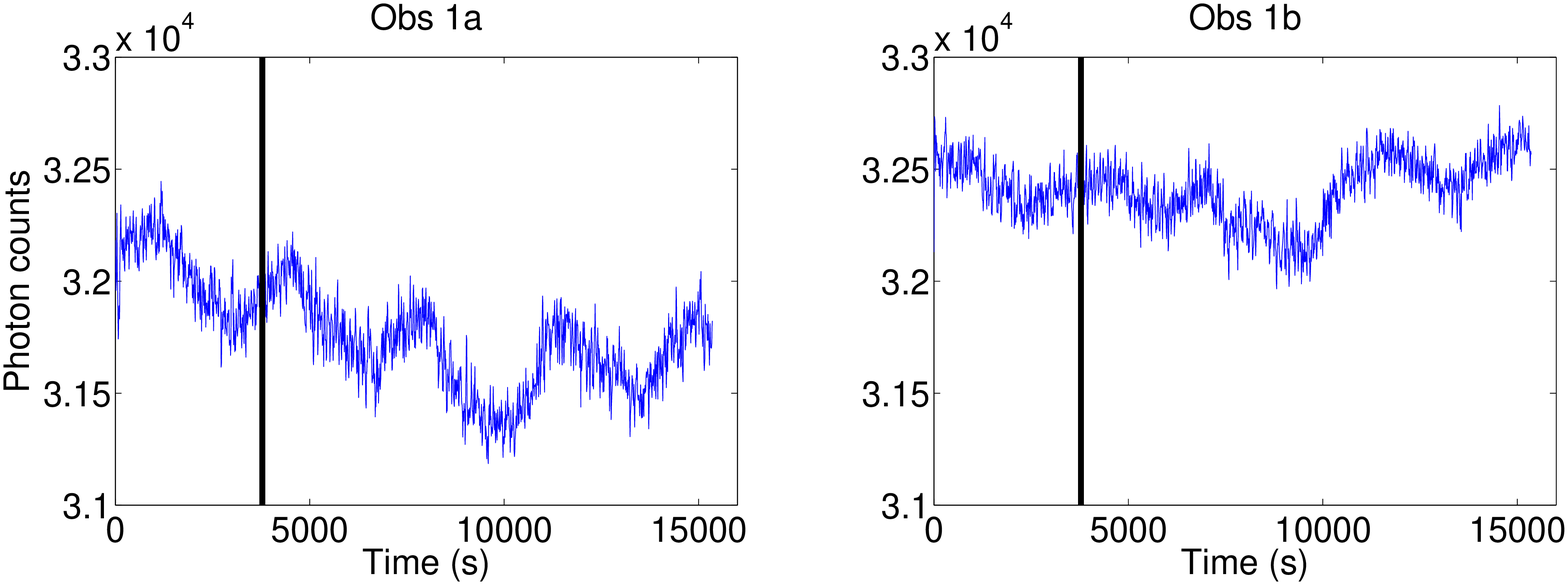}
\plotone{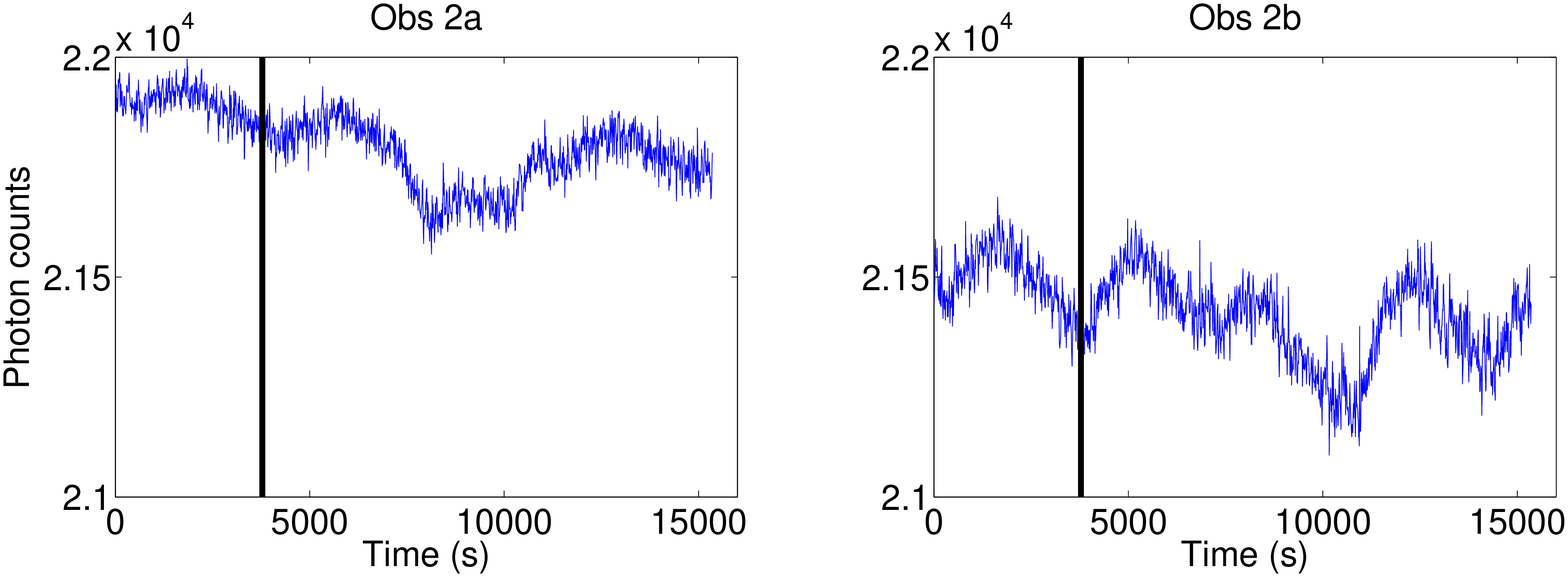}
\caption{Raw integral light-curves of the four primary transit observations. Data points on the left of black vertical lines have been discarded for the analysis. Note that the transit depth is comparable with the amplitude of systematics. \label{fig1}}
\end{figure}

Fig. \ref{fig1} reports the raw ``integral light-curves'' observed. The main systematic effect for IRAC channels 1 and 2 observations is an almost regular undulation with period $\sim$3000 s, so-called pixel-phase effect, because it depends on the relative position of the source centroid with respect to a pixel center \citep{faz04, mc06}. This effect is particularly difficult to detrend from these datasets because its timescale is similar to the transit duration, and its amplitude is comparable to the transit depth. Recently, a time dependence of the pixel-phase effect has been suggested \citep{ste10, bea11}.
\begin{figure}[!h]
%\epsscale{0.80}
\plotone{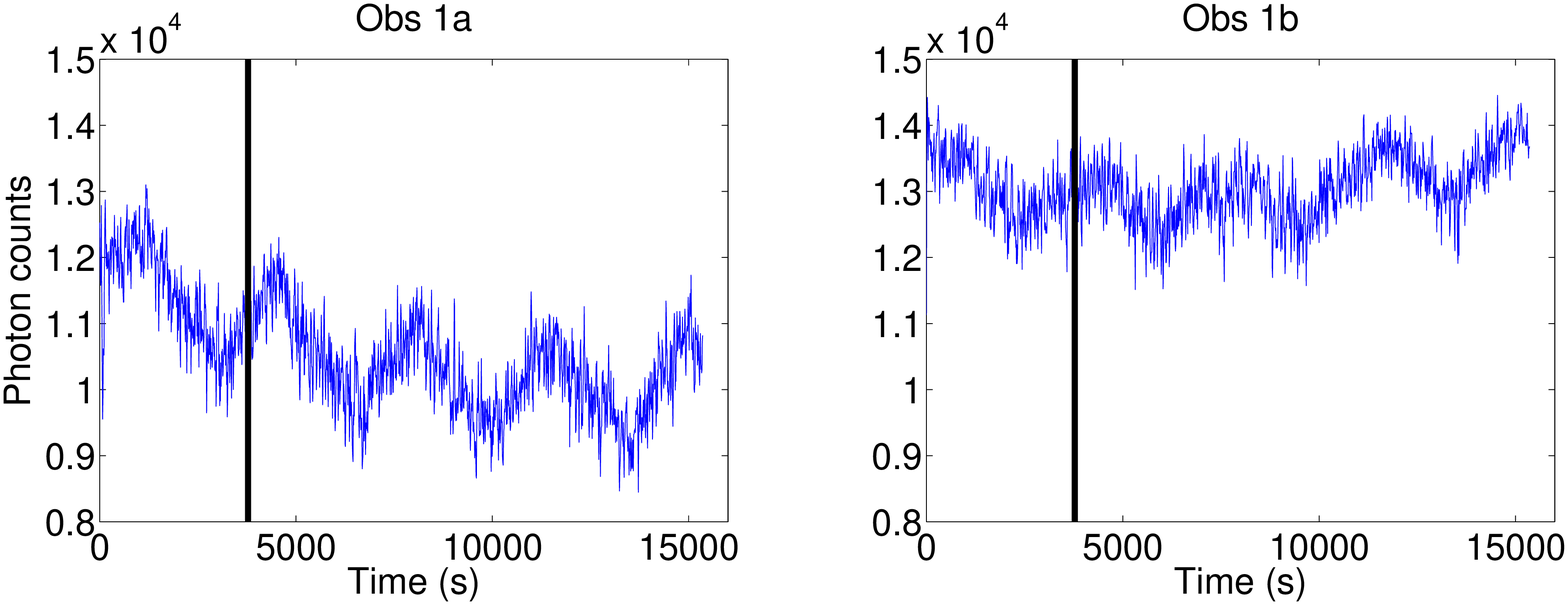}
\plotone{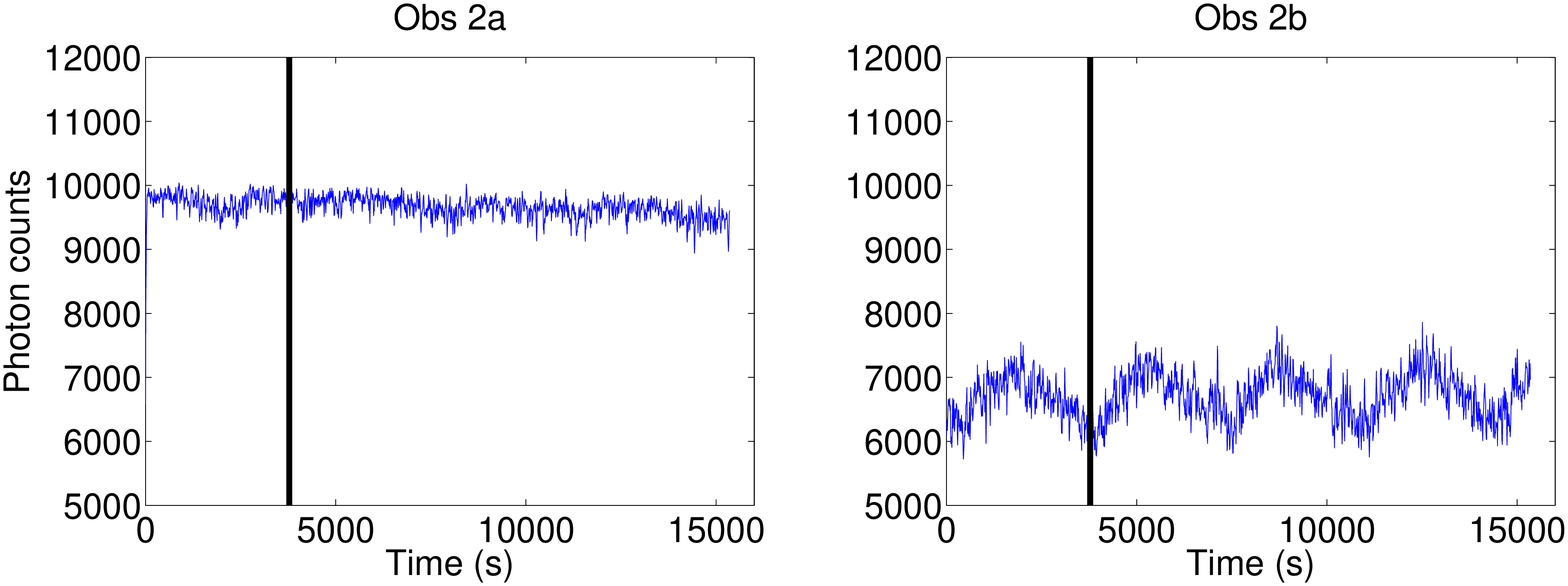}
\caption{Central pixel-light-curves of the four primary transit observations. Data points on the left of black vertical lines have been discarded for the analysis. Note that pixel-light-curves are dominated by systematics, and the transit signal is not visible by eye (but it is present,  as proven by ICA retrieval). \label{fig2}}
\end{figure}

If considering the whole datasets for detrending, our ICA algorithm is able to remove most of the non-flatness on the out-of-transits, and visibly improve the in-transit shapes, but some visible issues remain (see Fig. \ref{fig16}). We noted that results improve significantly if rejecting a number of data points from the beginning of each observation. It is discussed on a statistical basis in Sec. \ref{ssec:pcc}. A possible explanation is that first data points contain a long-tail variation until stabilization of the instruments  (\cite{faz04}, see also App. \ref{sec:app0added}); this is not a crucial point for the data analysis. In the rest of this paper we discuss the results obtained after rejecting the first 450 exposures from each observation, corresponding to $\sim$3780 s, for which the ICA performances are optimal. It is worth to point out that different choices (including no data rejections) give consistent results (within 1 $\sigma$), with larger or similar error bars.

\subsubsection{Limb darkening coefficients}
\label{ssec:ldc}

Tab. \ref{tab2} reports the quadratic limb darkening coefficients used at 3.6 and 4.5 $\mu$m IRAC passbands. Both the Atlas and Phoenix models are computed with T$_{eff}$ = 3680 K, $\log{g}$ = 4.78 \citep{tor09}, and solar abundances \citep{asp09}.
\begin{table}[!h]
\begin{center}
\caption{Quadratic limb darkening coefficients computed by Atlas and Phoenix stellar models for IRAC 3.6 and 4.5 $\mu$m bands. \label{tab2}}
\begin{tabular}{ccc}
\tableline\tableline
Atlas & 3.6 $\mu$m & 4.5 $\mu$m \\
\tableline
$\gamma_1$ & 5.489 $\times$ 10$^{-2}$ & 1.331 $\times$ 10$^{-2}$\\
$\gamma_2$ & 3.0653 $\times$ 10$^{-1}$ & 2.8396 $\times$ 10$^{-1}$\\
\tableline
Phoenix & 3.6 $\mu$m & 4.5 $\mu$m \\
\tableline
$\gamma_1$ & 3.87 $\times$ 10$^{-3}$ & 3.27 $\times$ 10$^{-3}$\\
$\gamma_2$ & 2.3615 $\times$ 10$^{-1}$ & 1.8193 $\times$ 10$^{-1}$\\
\tableline
\end{tabular}
%% Any table notes must follow the \end{tabular} command.
\end{center}
\end{table}

\section{Results}

\subsection{Tests of pixel-phase correlations}
\label{ssec:pcc}

\begin{figure}[!h]
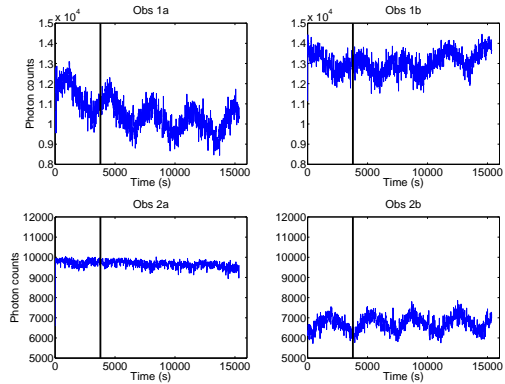

%\epsscale{0.70}
\plotone{f2a.eps}
\plotone{f2b.eps}
%\plotone{pixelphase_obs1a.eps}
%\plotone{pixelphase_obs1b.eps}
%\plotone{pixelphase_obs2a.eps}
%\plotone{pixelphase_obs2b.eps}
\caption{Time series of the pixel-phase values for the four observations. Data points on the left of black vertical lines have been discarded for the analysis; dashed green lines delimit the ends of pre-transits and the begins of post-transits; dashed red lines delimit the in-transits. \label{fig3}}
\end{figure}

\begin{figure}[!h]
%\epsscale{0.70}
%\plotone{pearson_all.eps}
\plotone{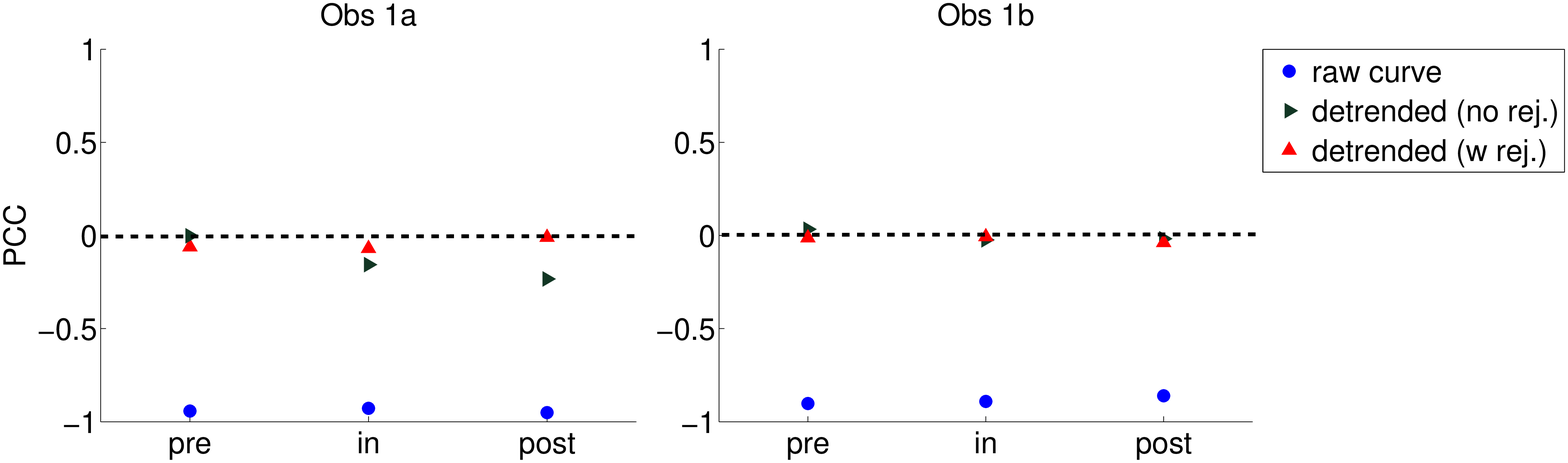}
\plotone{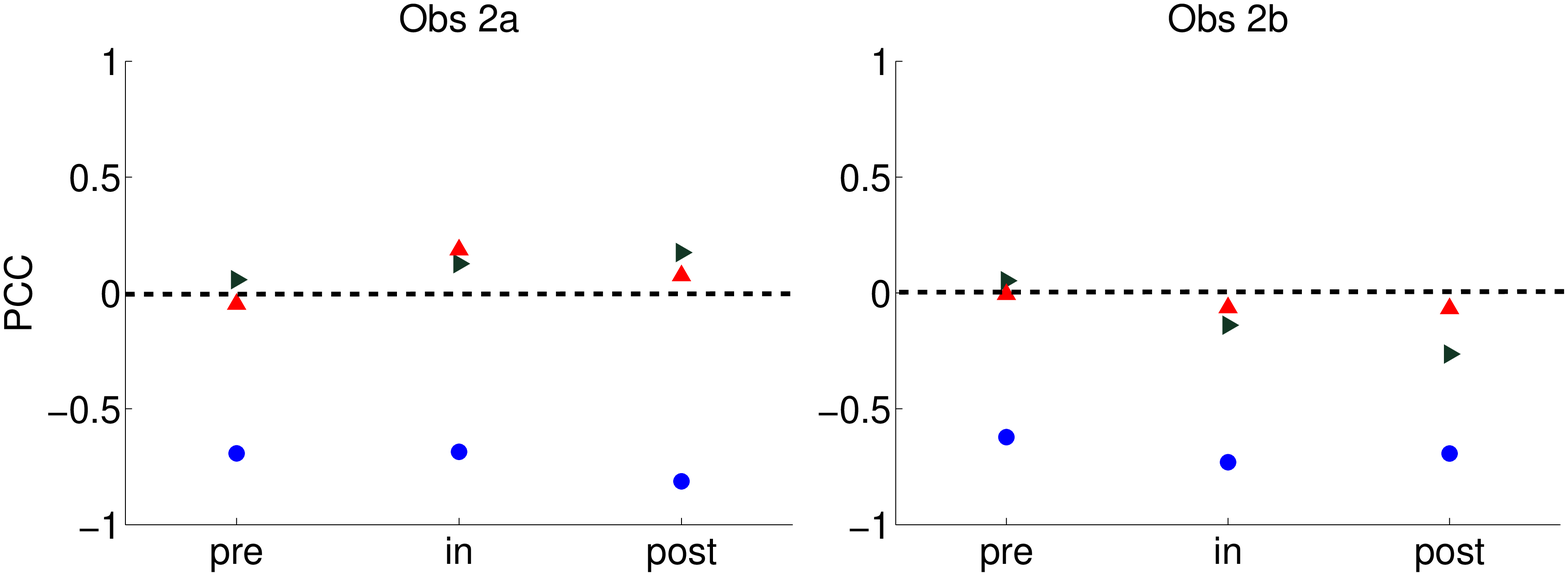}
\caption{PCCs between fluxes and pixel-phases for pre-, in-, and post-transits of the four light-curves; (blu circles) raw data, (green rightwards triangles) ICA detrended data with no rejections, (red upwards triangles) ICA detrended data after rejecting the first 450 points. \label{fig4}}
\end{figure}

\begin{figure}[!h]
%\epsscale{0.70}
\plotone{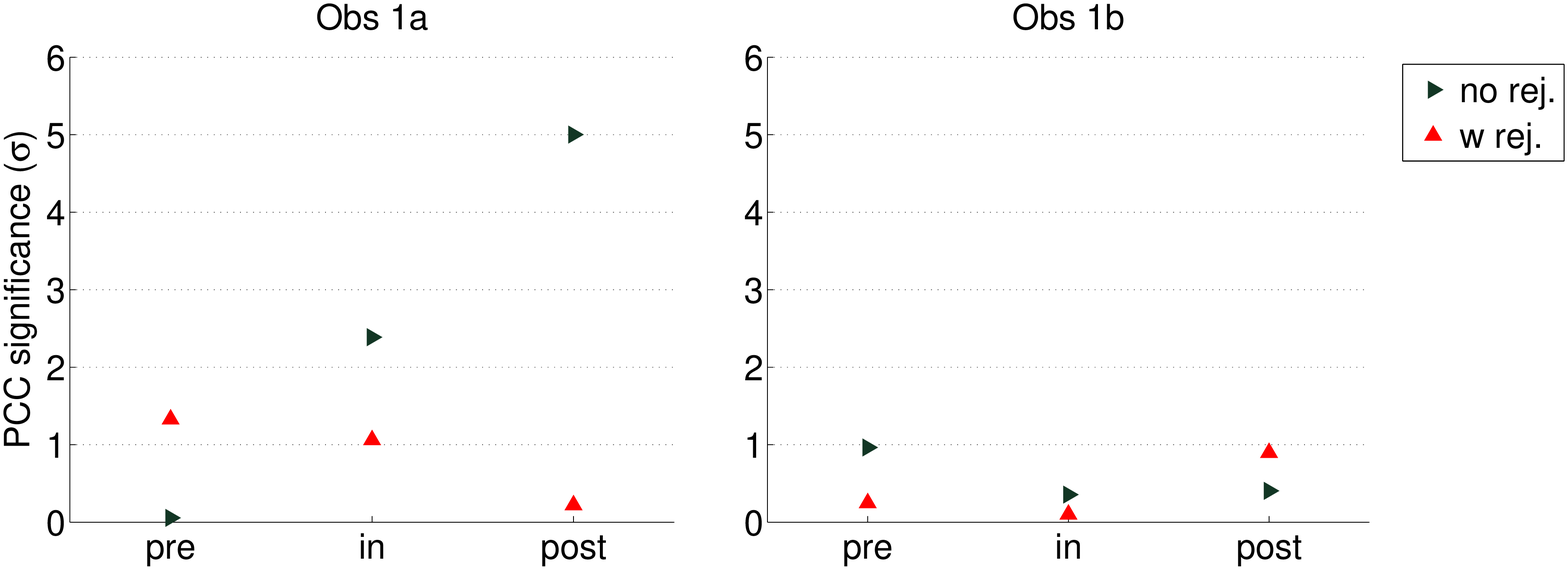}
\plotone{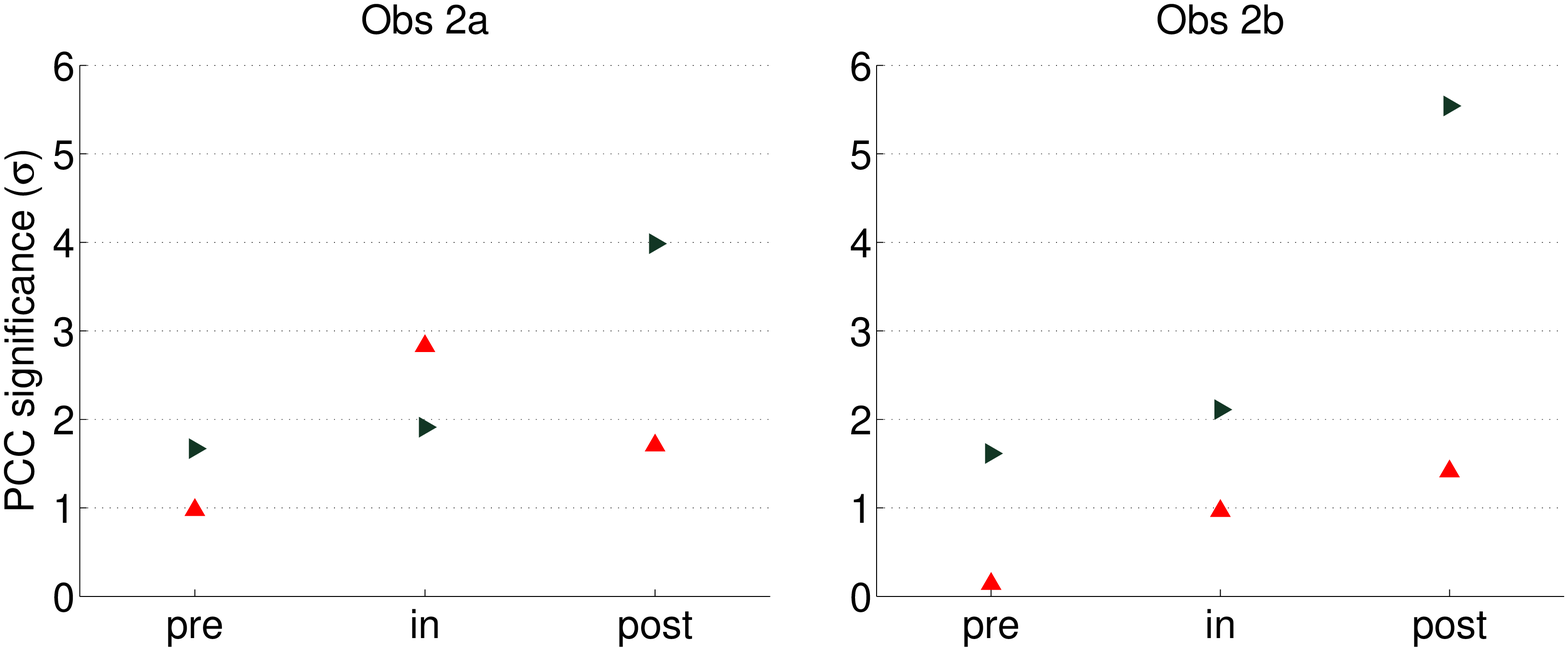}
\caption{Significance level of correlation between fluxes and pixel-phases for the four observations; (green rightwards triangles) ICA detrended data with no rejections, (red upwards triangles) ICA detrended data after rejecting the first 450 points. \label{fig5}}
\end{figure}

To investigate the effectiveness of the data detrending we measure the correlations of the signals with the pixel-phase position, before and after the corrections. We refer to the Pearson product-moment correlation coefficient (PCC), defined as:
\begin{equation}
PCC = \frac{ cov(X,Y)}{ \sigma_X \sigma_Y}
\end{equation}
where $cov(X,Y)$ is the covariance of the signals $X$ and $Y$, $\sigma_X$ and $\sigma_Y$ are the standard deviations. In this context, $X$ and $Y$ are temporal series of fluxes and pixel-phases. The PCCs are measured over three intervals, i.e. pre-, in-, and post-transit, where the astrophysical signals are expected to be almost flat \footnote{We used the following definitions: pre-transit ($\phi <$ -0.0082); in-transit (-0.00433 $< \phi < $0.00416); post-transit ($\phi > $0.0082). These have been decided so that all the transit models obtained during the analysis, modified with no limb darkening, are flat in these three intervals. We checked that other reasonable choices of the limits do not affect this analysis.}. In general -1$\le$PCC$\le$+1, where +1 is total positive correlation, -1 is total negative correlation, and 0 is no correlation. Fig. \ref{fig3} reports the temporal series of pixel-phases. Fig. \ref{fig4} reports the values of the PCCs measured on the pre-, in-, and post-transit for each observation, uncorrected, corrected without and with pre-transit truncation.
The original data are strongly anticorrelated with the pixel-phase, with PCC$\lesssim$-0.9 for channel 1, and PCC$\sim$-0.7 for channel 2. After the ICA detrending including all the data, these correlations are significatively reduced ($|$PCC$| <$0.3). If we remove the first 450 data points, the ICA detrending generally performs significantly better ($|$PCC$| \sim$10$^{-3}$-7$\times$10$^{-2}$). Fig. \ref{fig5} reports the level of significance of the residual correlations in the detrended data, calculated with a permutation test. When we reject the first 450 data points, the residual correlations in the detrended data are below 1.5 $\sigma$, except for Obs. 2a, for which the residual correlation is higher in any case.  The residual correlations without the cut of the first 450 data points are larger, i.e. $>$4 $\sigma$ in the post transit, with the exception of Obs. 1b, for which the residual correlations are below 1 $\sigma$ in any case.

\subsection{Fitting $p$, $a_0$ and $i$}
\label{sec:starting_fits}

\begin{figure}[!h]
%\epsscale{0.70}
\plotone{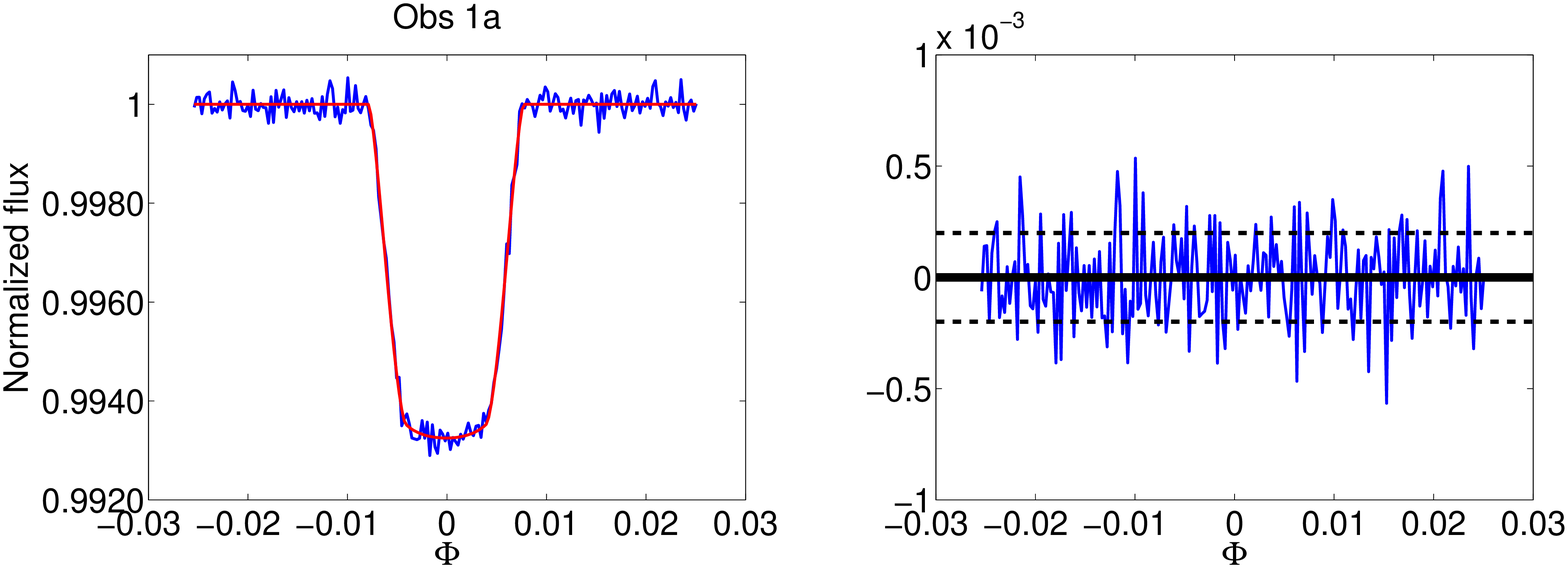}
\plotone{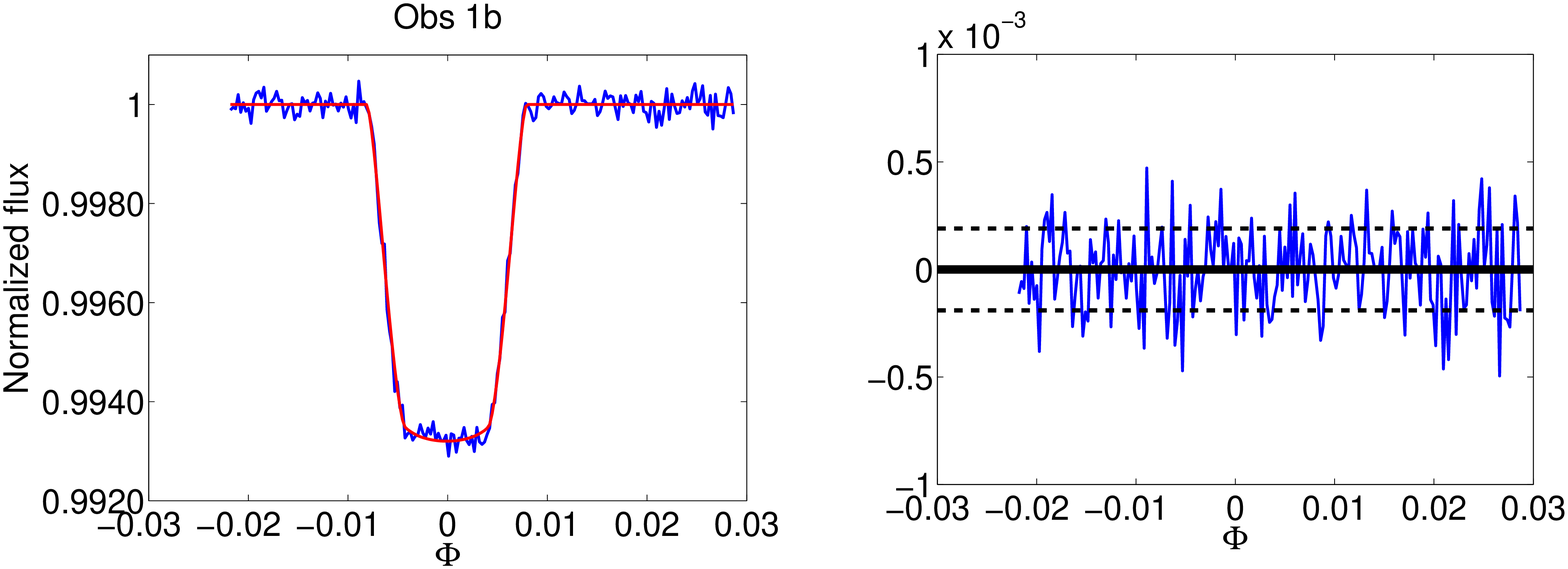}
\plotone{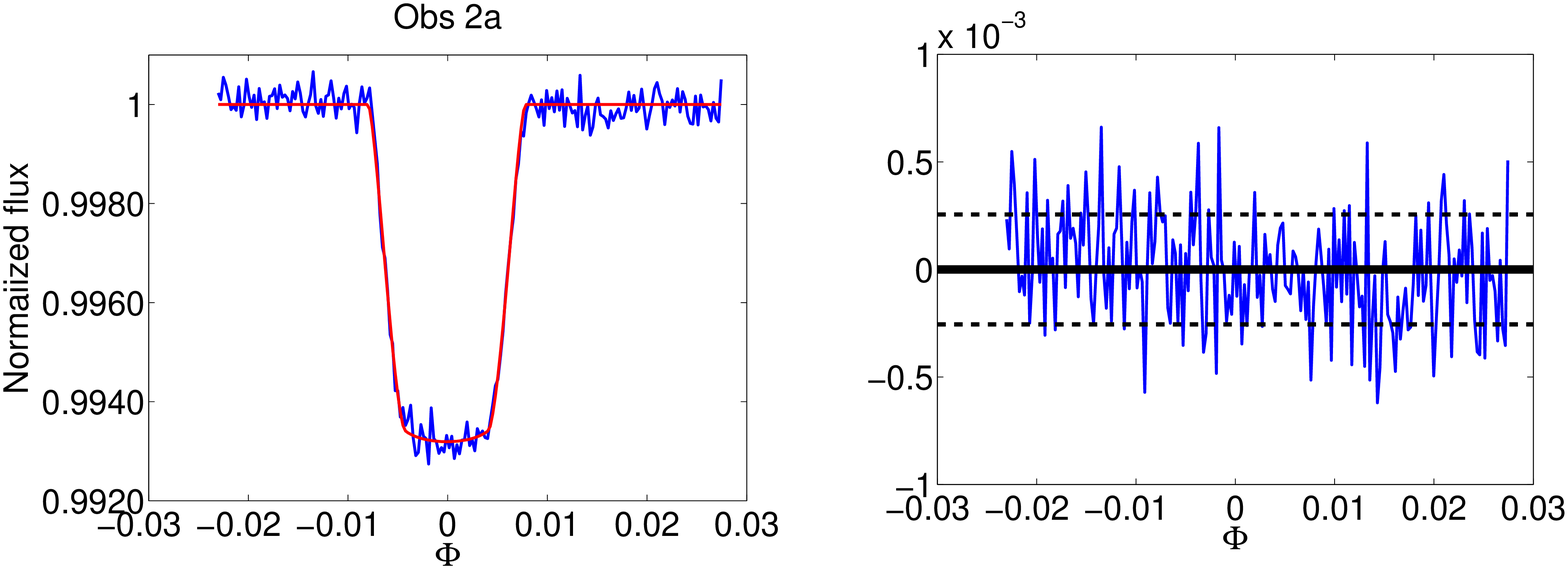}
\plotone{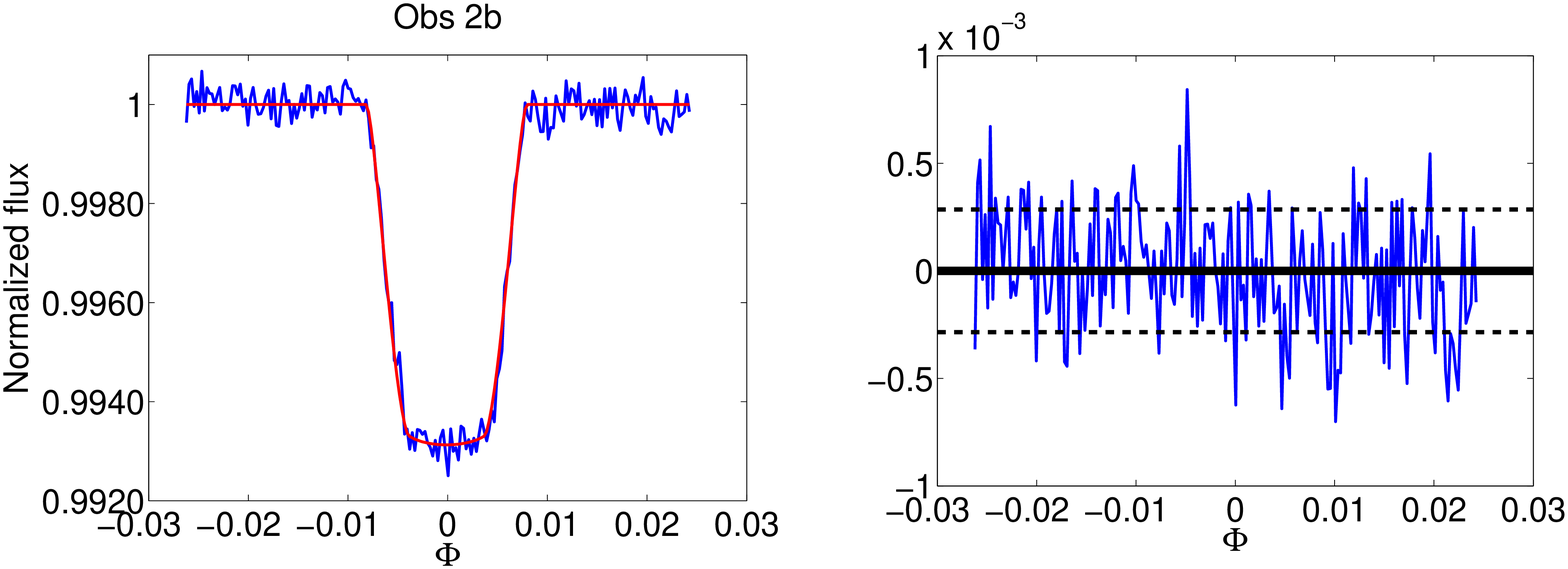}
\caption{Left panels: (blue) detrended light-curves for the four observations with (red) best transit models overplotted, binned over 7 points; best transit models are calculated with $p$, $a_0$, and $i$ as free parameters, and Phoenix quadratic limb darkening coefficients (see Sec. \ref{ssec:ldc}). Right panels: Residuals between detrended light-curves and best transit models; black horizontal dashed lines indicate the standard deviations of residuals. \label{fig6}}
\end{figure}

Fig. \ref{fig6} reports the detrended light-curves, binned over 7 points, with the relative best transit models, and the residuals. The transit models in Fig. \ref{fig6} are computed with $\gamma_1$ and $\gamma_2$ Phoenix coefficients. Analogous transit models computed with $\gamma_1$ and $\gamma_2$ Atlas coefficients are very similar, with average standard deviations $\lesssim$1.9$\times$10$^{-5}$, and maximum discrepacies $\lesssim$10$^{-4}$. Discrepancies between the transit models and the detrended light-curves are at the level $\sim$2.0$\times$10$^{-4}$ for IRAC channel 1, and $\sim$2.6$-$2.9$\times$10$^{-4}$ for IRAC channel 2, therefore it is not possible to distinguish between Atlas and Phoenix models from the data.
Best parameter results and error bars are reported in Fig. \ref{fig7}, in Tab. \ref{tab4} and \ref{tab5}.
%, together with the sampled $\sigma_0$ values. Note that $\sigma_0$ values are comparable with the standard deviations of residuals.
Atlas and Phoenix stellar models lead to two systematically different parameter sets, but within the error bars. All the parameters from different observations are comparable within 1$\sigma$, even neglecting the detrending errors ($\sigma_{ICA}$), except the transit durations for Obs 1a and 1b. This is discussed in the following sections.

\subsubsection{Combining observations}
\label{sec:combined_fits}

We performed two couples of simultaneos fits, one for the 3.6 $\mu$m and one for the 4.5 $\mu$m light-curves, with Atlas and Phoenix limb darkening coefficients, assuming common orbital parameters ($a_0$ and $i$), and potentially different transit depths ($p$), in order to cancel the effects of parameter intercorrelations. The assumption that orbital parameters are the same during each observation is very reliable, because they are sparse over a short period of time (less than 1 month, 9 planetary orbital periods), so that variations due to relativistic effects, external perturbers or tidal effects, would be very small compared to the error bars \citep{alo08, jor08, pal08}.

The results of these combined fits are reported in Fig. \ref{fig8}, \ref{fig9}, in Tab. \ref{tab6} and \ref{tab7}.
The 4.5 $\mu$m transit depths become identical, with an intermediate value between the two determined with separate fits; the 3.6 $\mu$m transit depths slightly diverges, but their separation is still less than 1$\sigma$. The standard deviations of residuals between the detrended light-curves and the transit models increase of $\sim$2$-$3$\times$10$^{-6}$ for Obs 2a and 2b (negligible), and of $\sim$7$-$8$\times$10$^{-6}$ for Obs 1a and 1b (comparable with the $\sigma_0$ uncertainties). The assumption of common orbital parameters for Obs 2a and 2b may be valid, being the consequent transit models as good as the individually fitted ones. Being the transit depths also identical, the two light-curves are very well approximated by the same transit model. The original discrepancies between the two sets of transit parameters were enlarged by their intercorrelations. The same assumption for Obs 1a and 1b lead to worse transit models and more divergent transit depths, but, in both cases, not dramatically. 

\subsection{Timing variations}
\label{ssec:TTV}

We performed transit model-fits with a free phase-shift in addition to $p$, $a_0$, and $i$, in order to investigate the effect of possible timing variations. Fig. \ref{fig10} reports the time-shifts obtained. No evidence of timing variation have been detected, with upper limits $<$30 s. Both Atlas and Phoenix stellar models lead to the same shifts. Other parameter estimates are not affected.

\begin{figure}[!h]
%\epsscale{0.49}
\plotone{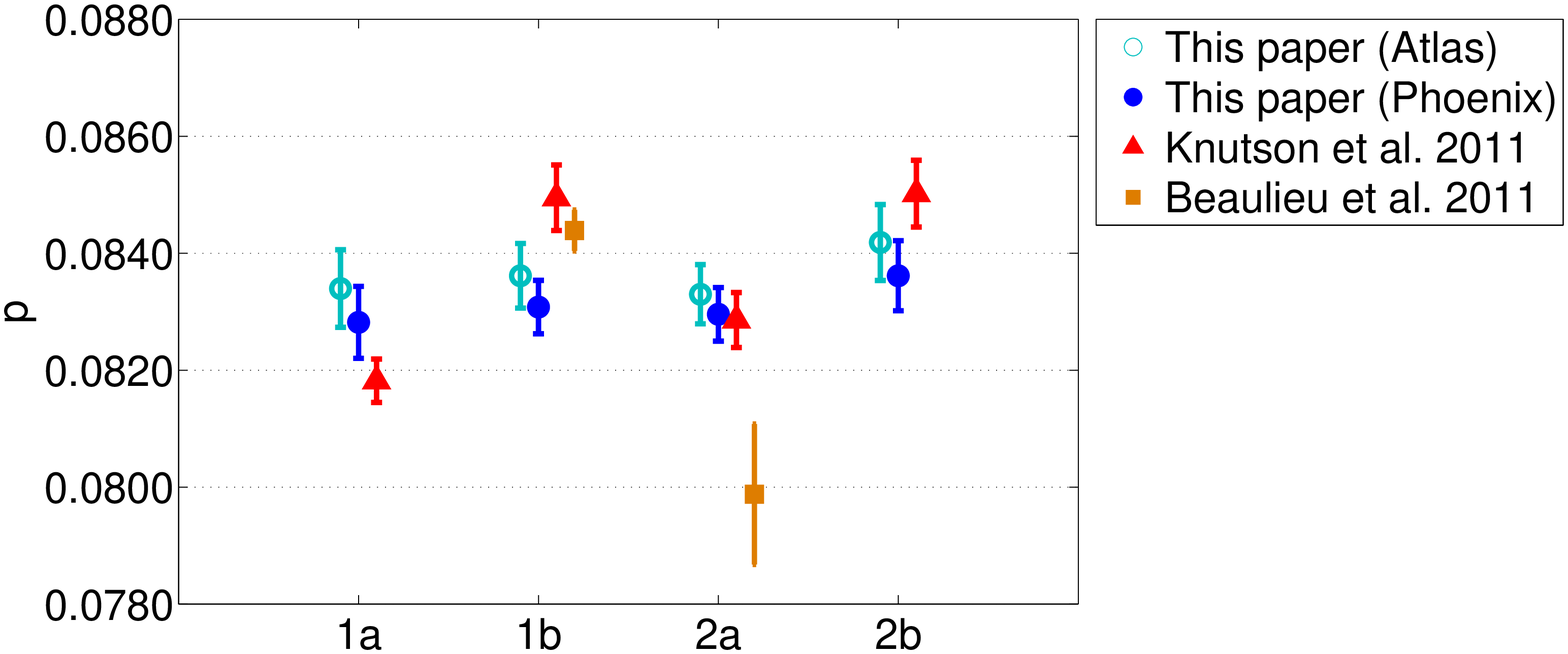}
\plotone{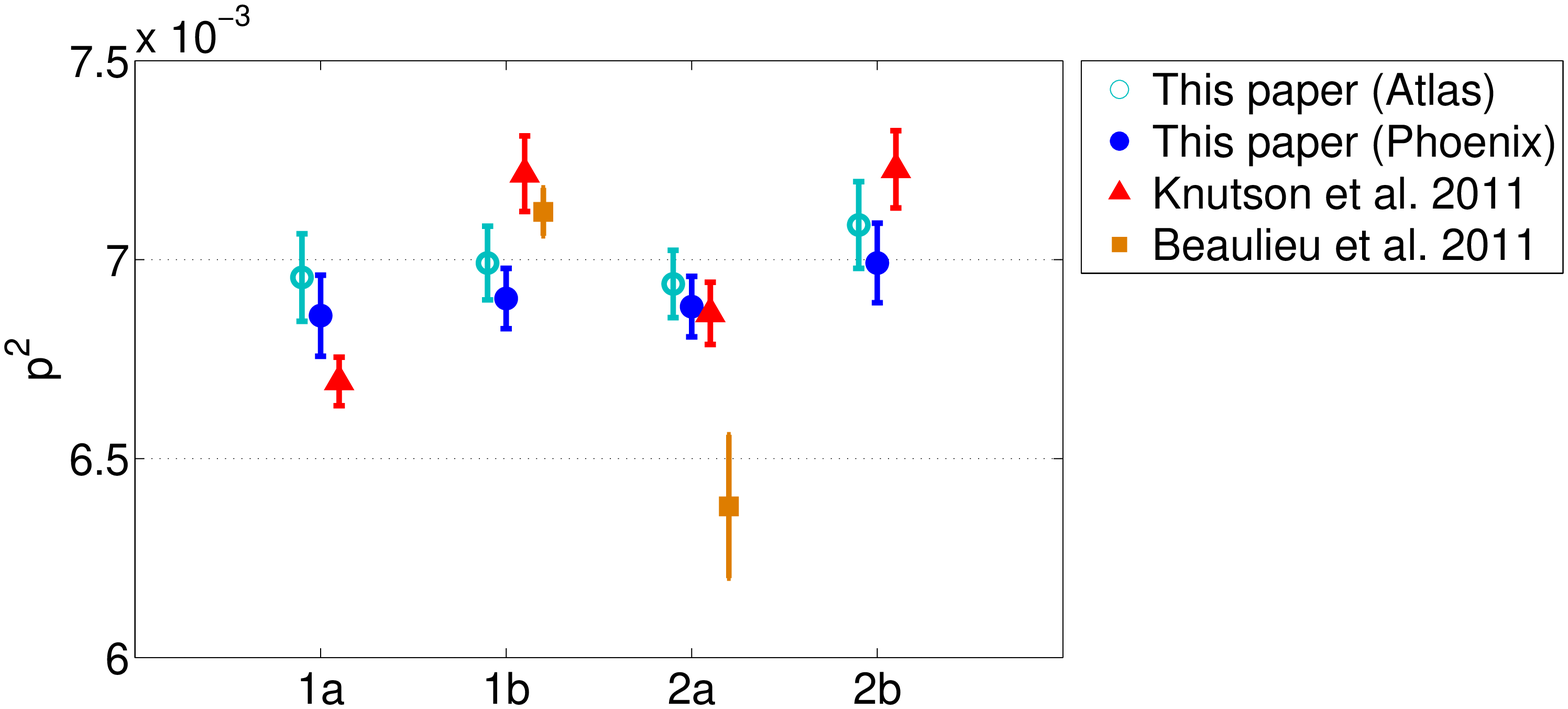}
\plotone{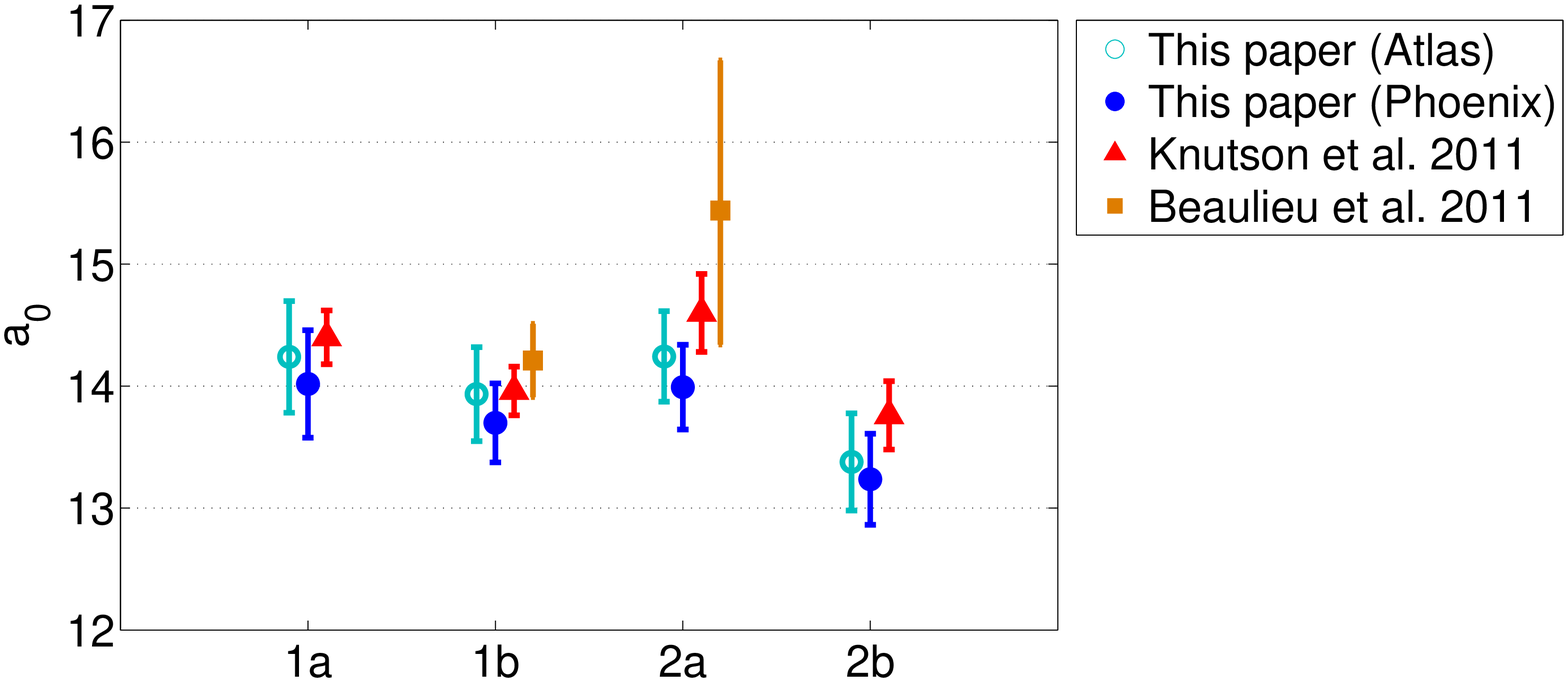}
\plotone{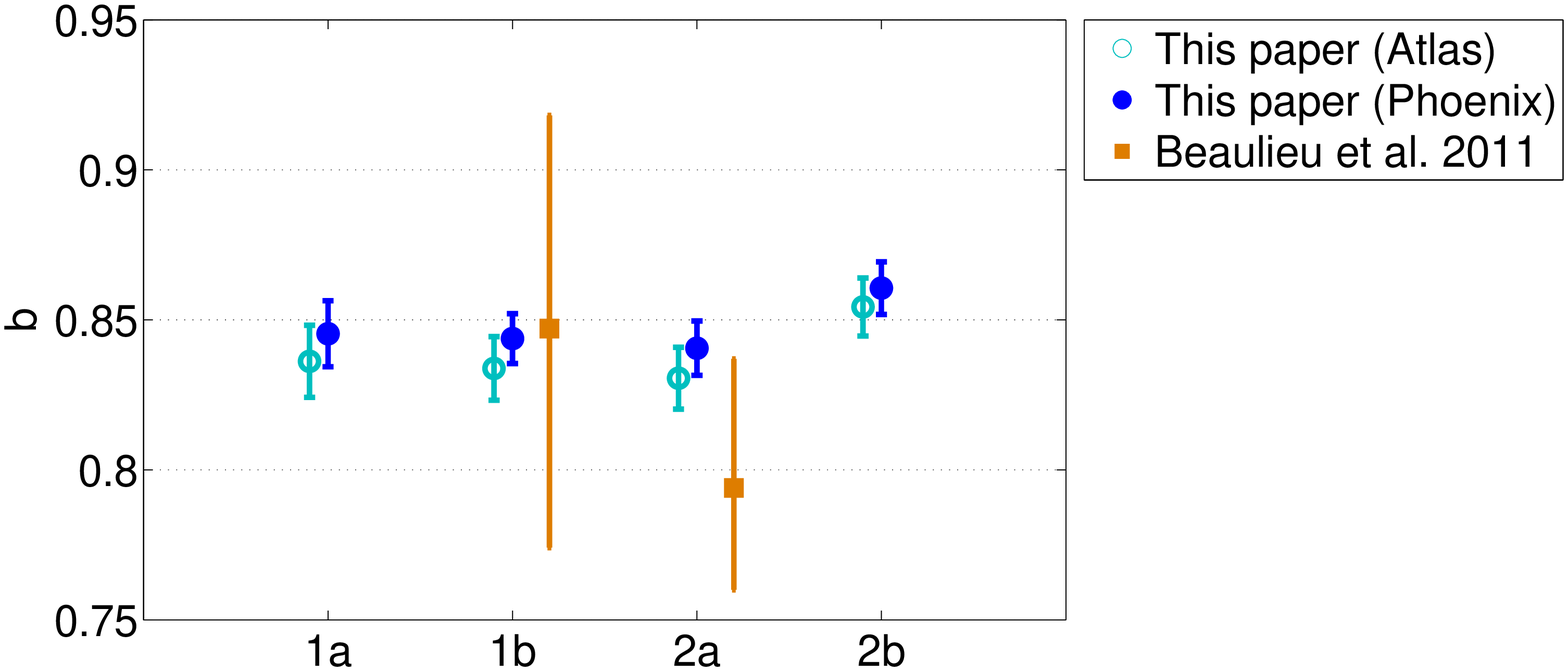}
\plotone{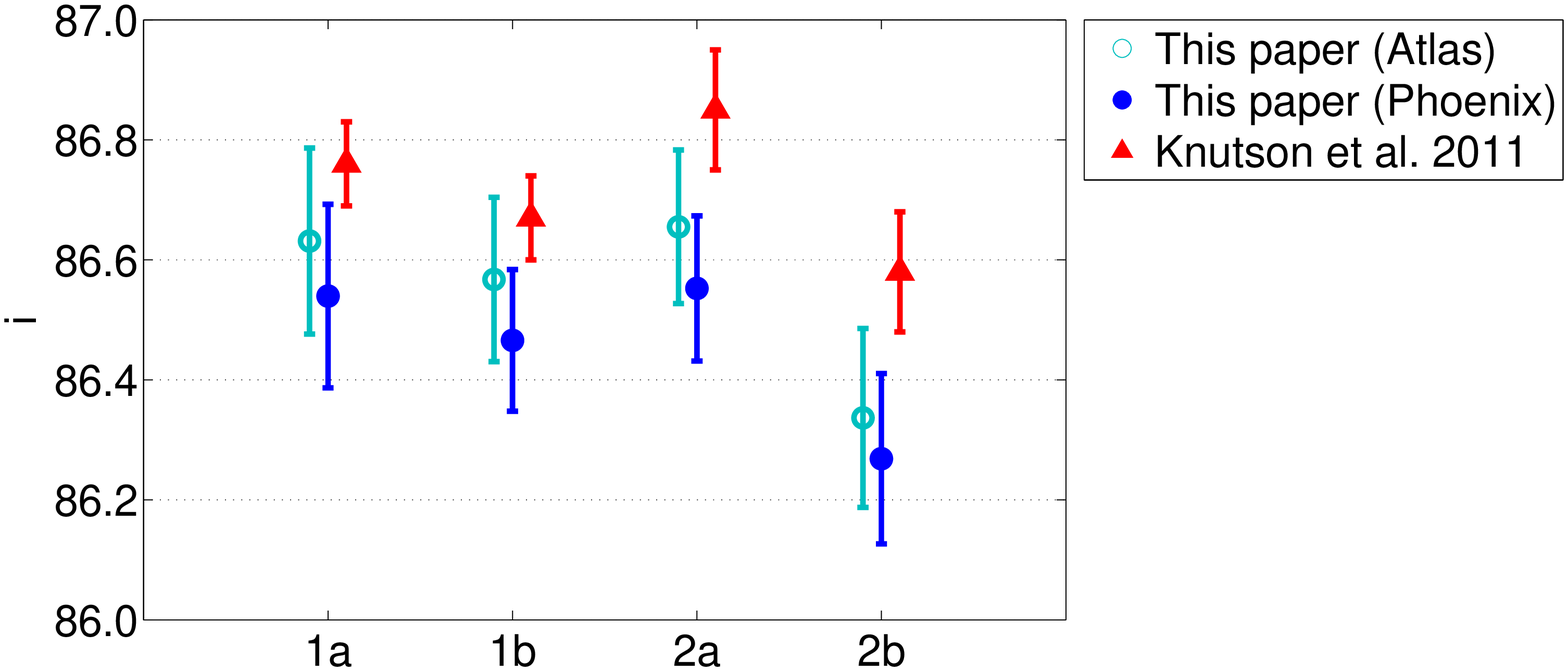}
\plotone{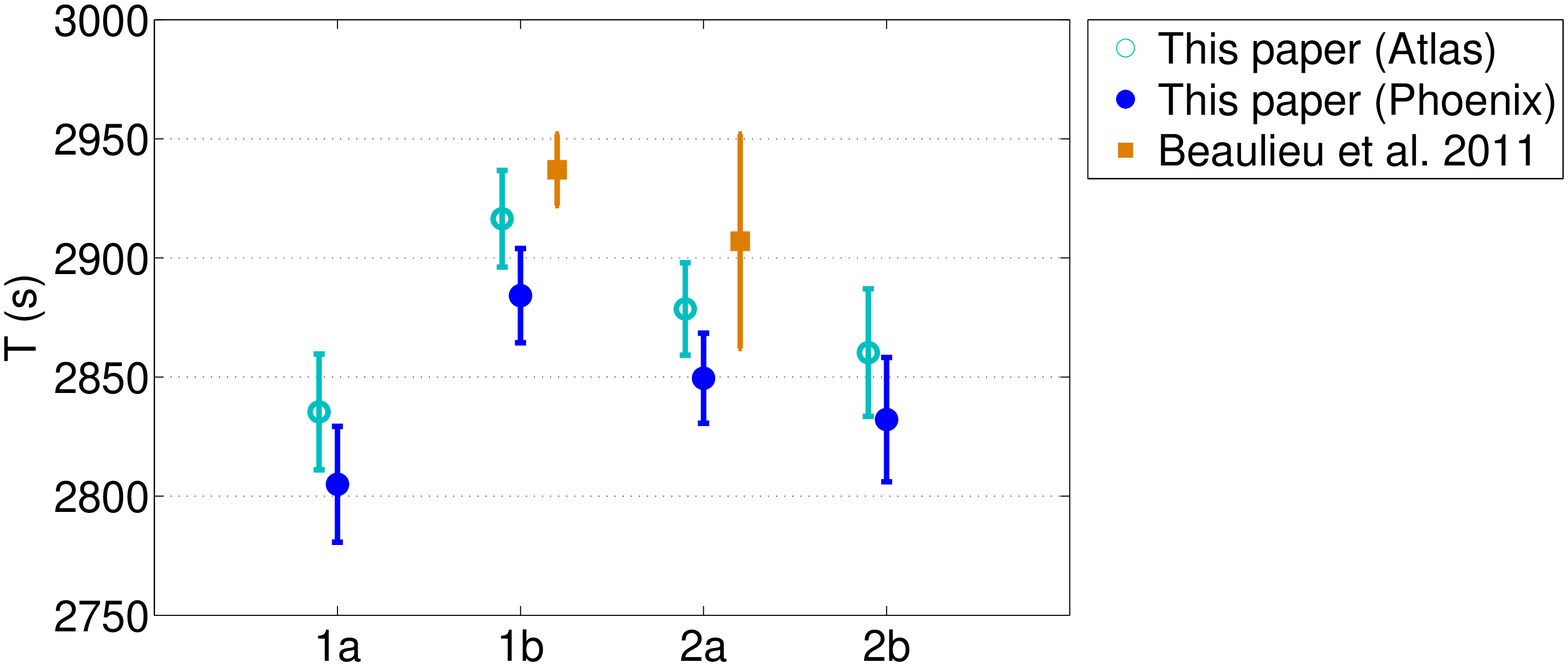}
\caption{From top to bottom: Comparisons of the parameters $p$, $a_0$, and $i$ (left side), $p^2$, $b$, and $T$ (right side), obtained in this paper with Atlas stellar model (cyan, empty circles), Phoenix stellar model (blue, full circles), in \cite{knu11} (red triangles), and in \cite{bea11} (yellow squares). \label{fig7}}
\end{figure}

\begin{figure}[!h]
%\epsscale{0.49}
\plotone{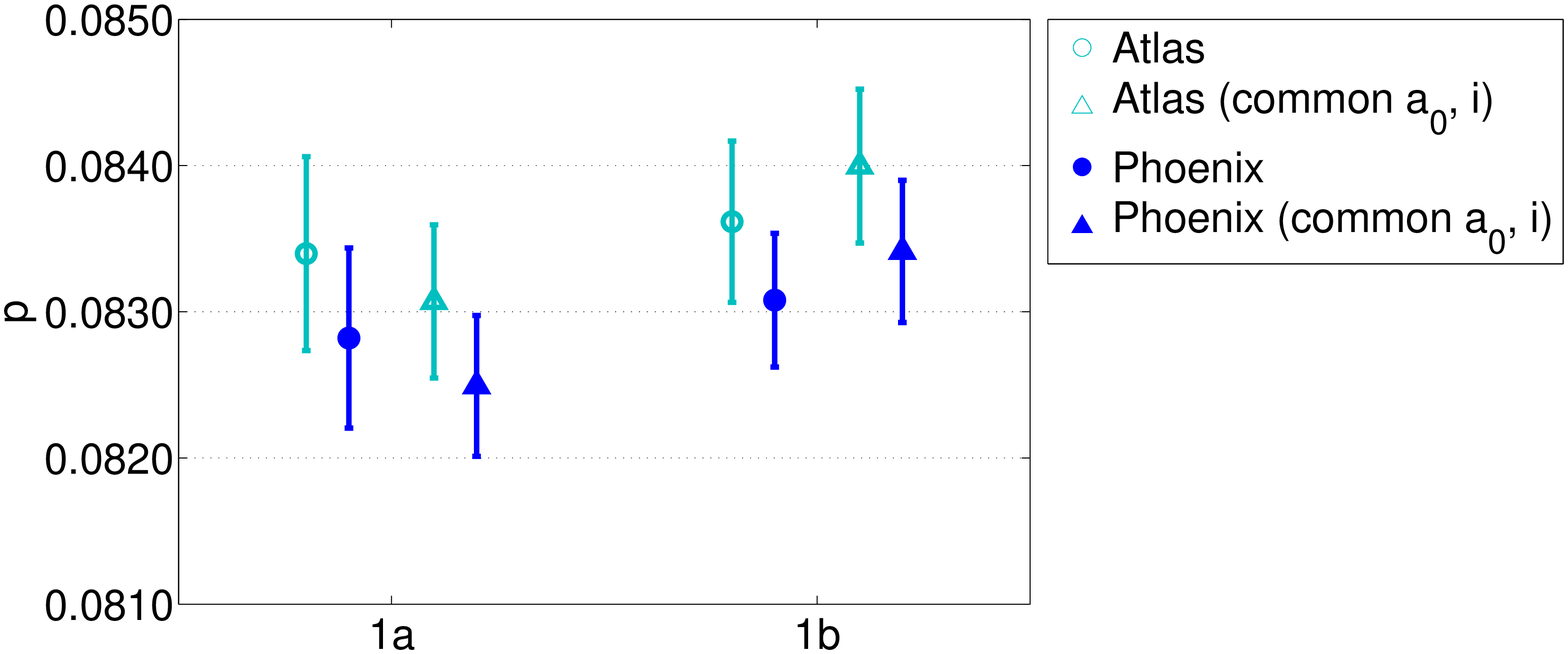}
\plotone{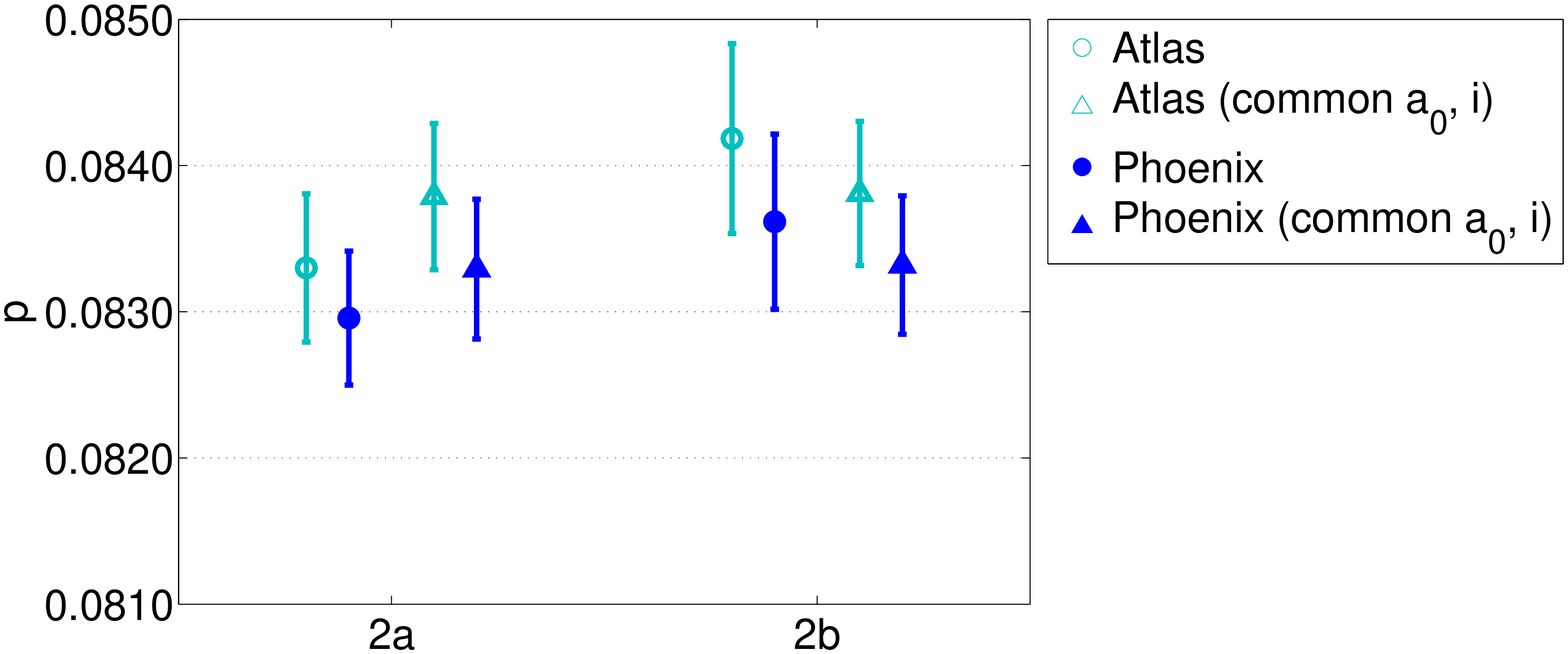}
\plotone{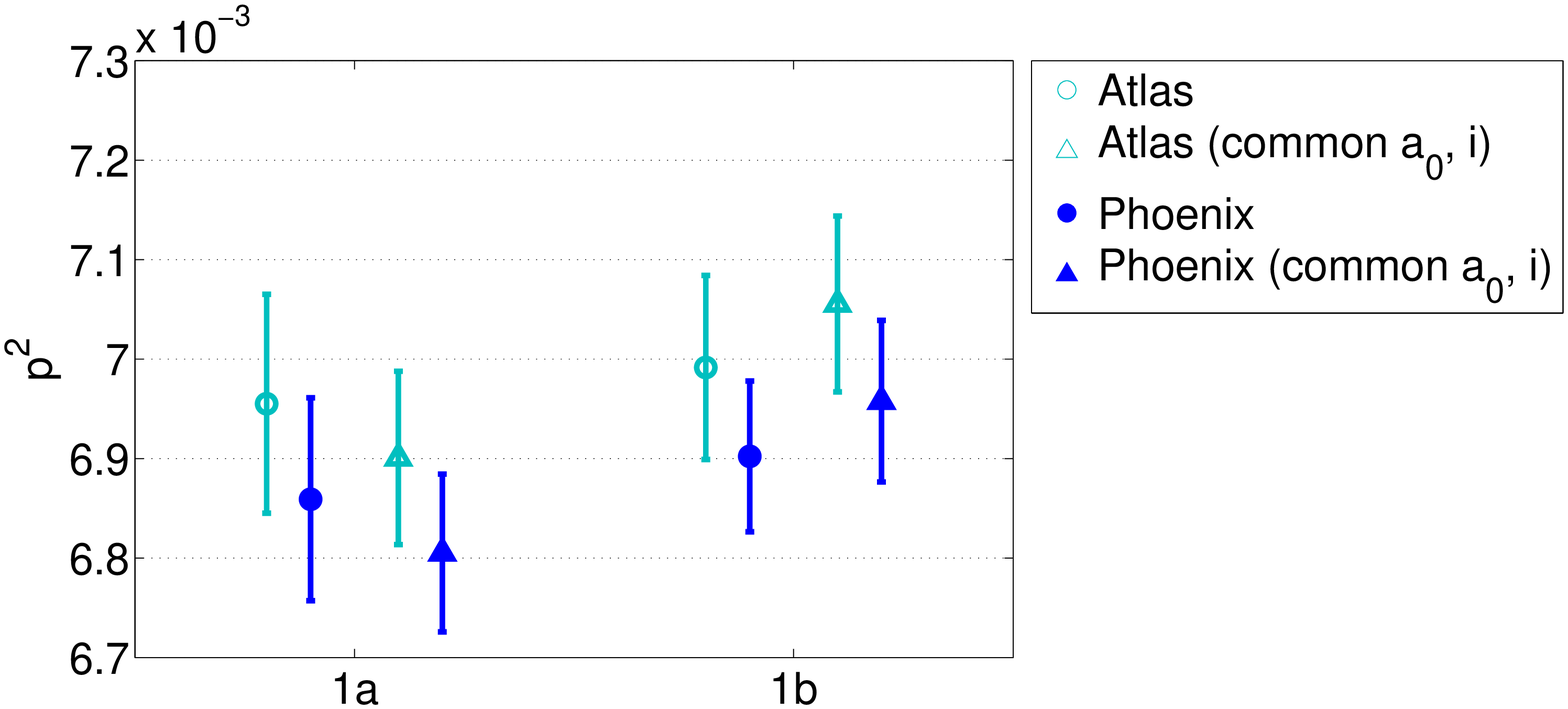}
\plotone{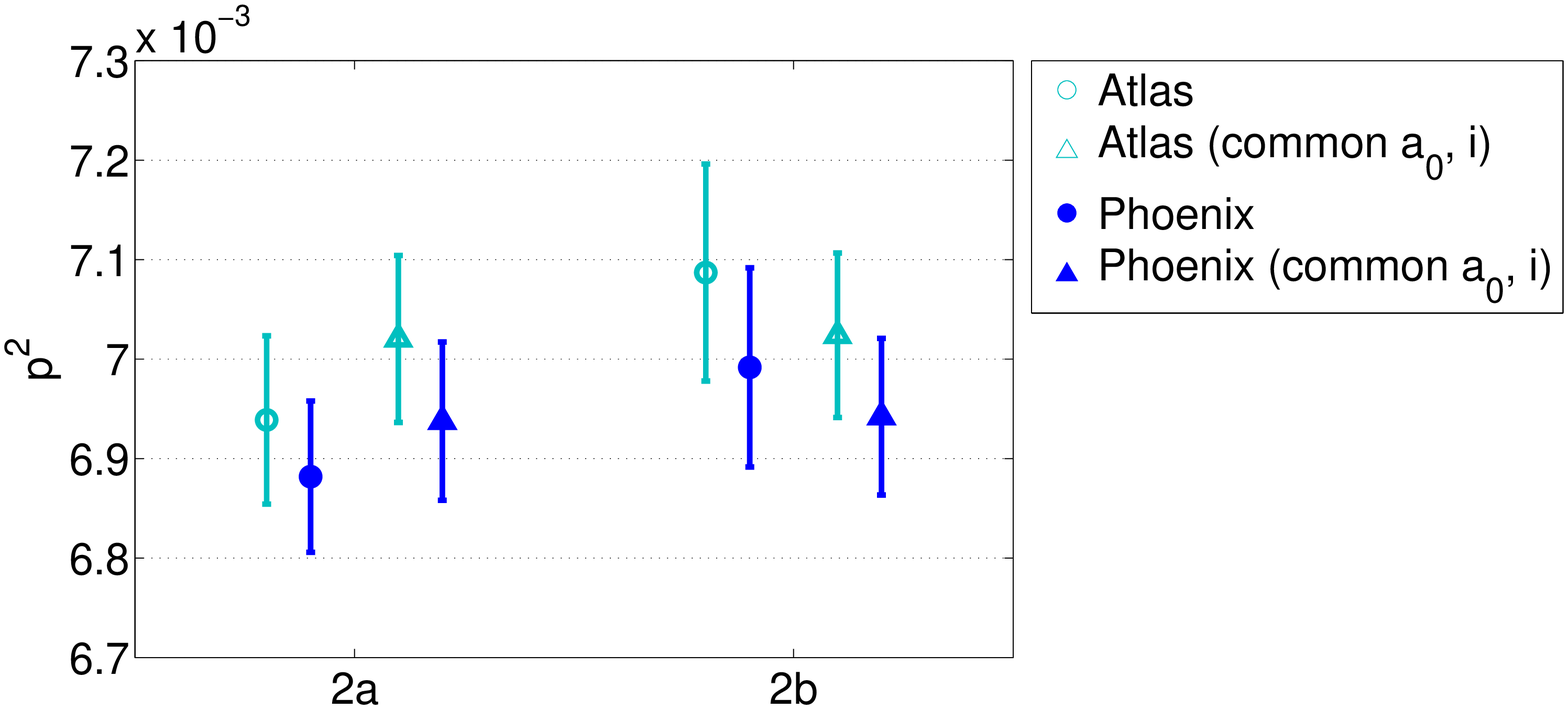}
\caption{Comparisons of the parameters $p$ and $p^2$, obtained in this paper, with common orbital parameters for observations at the same wavelength, and Atlas or Phoenix stellar models. \label{fig8}}
\end{figure}
\begin{figure}[!h]
%\epsscale{0.49}
\plotone{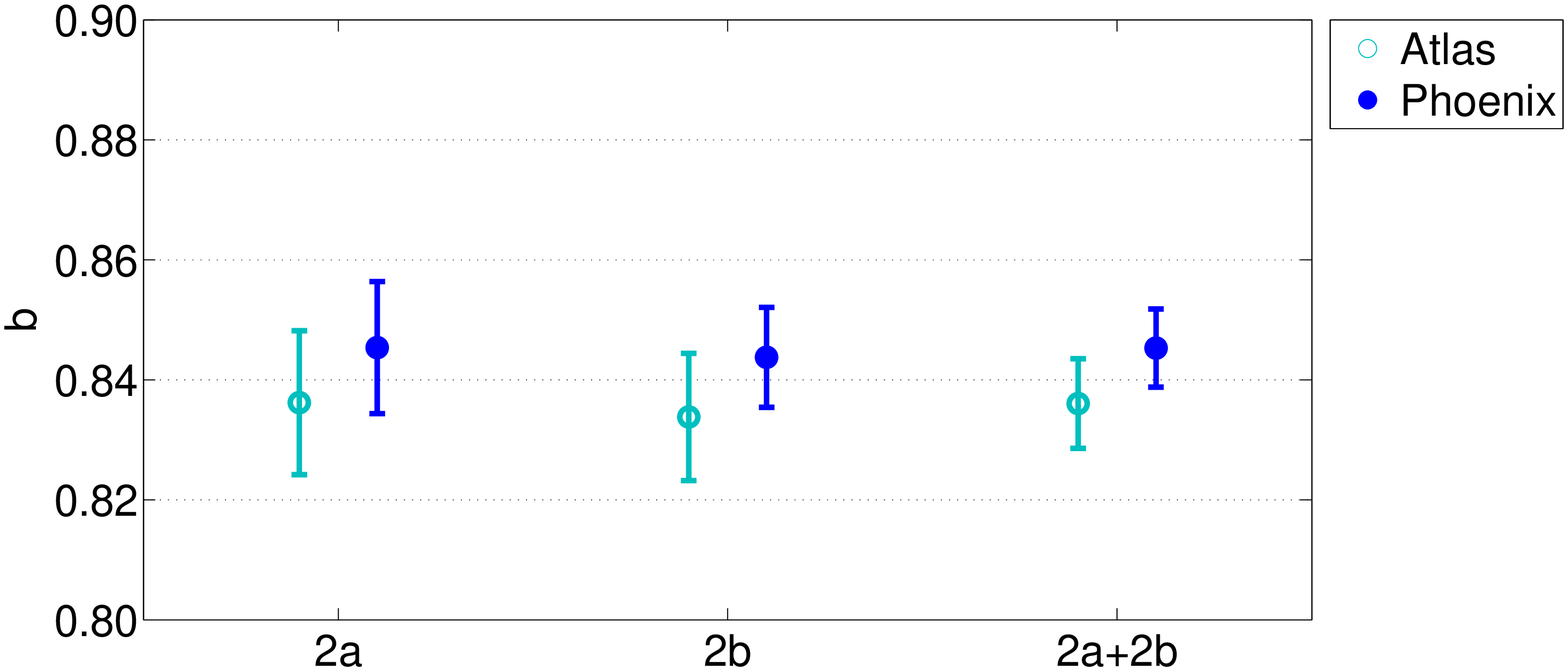}
\plotone{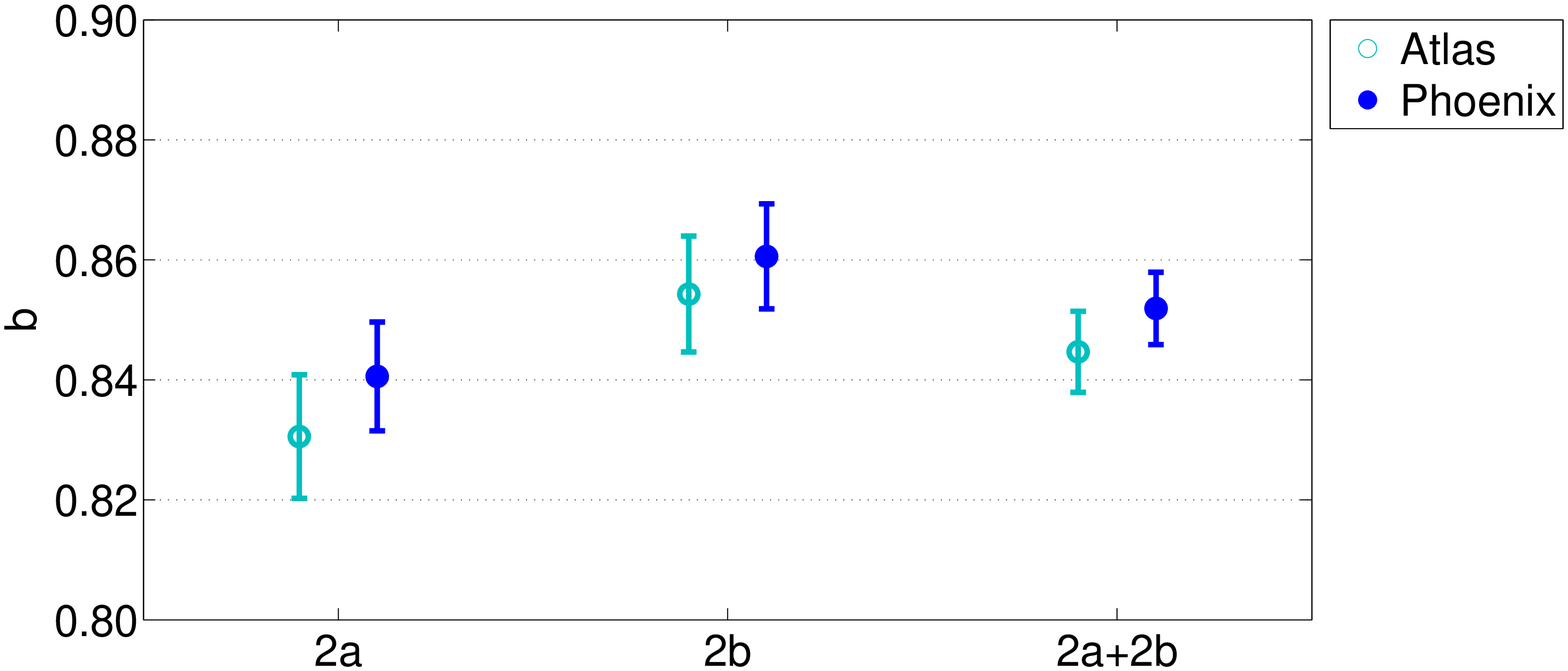}
\plotone{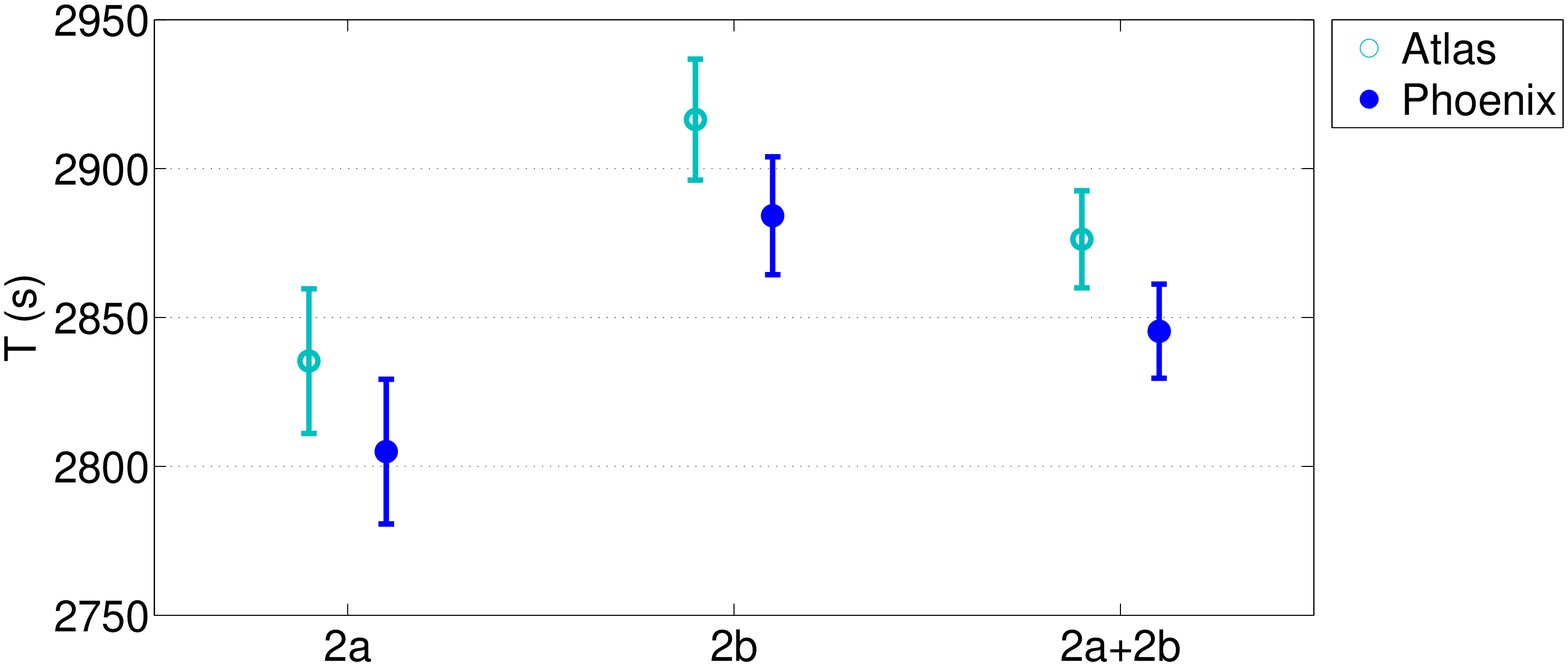}
\plotone{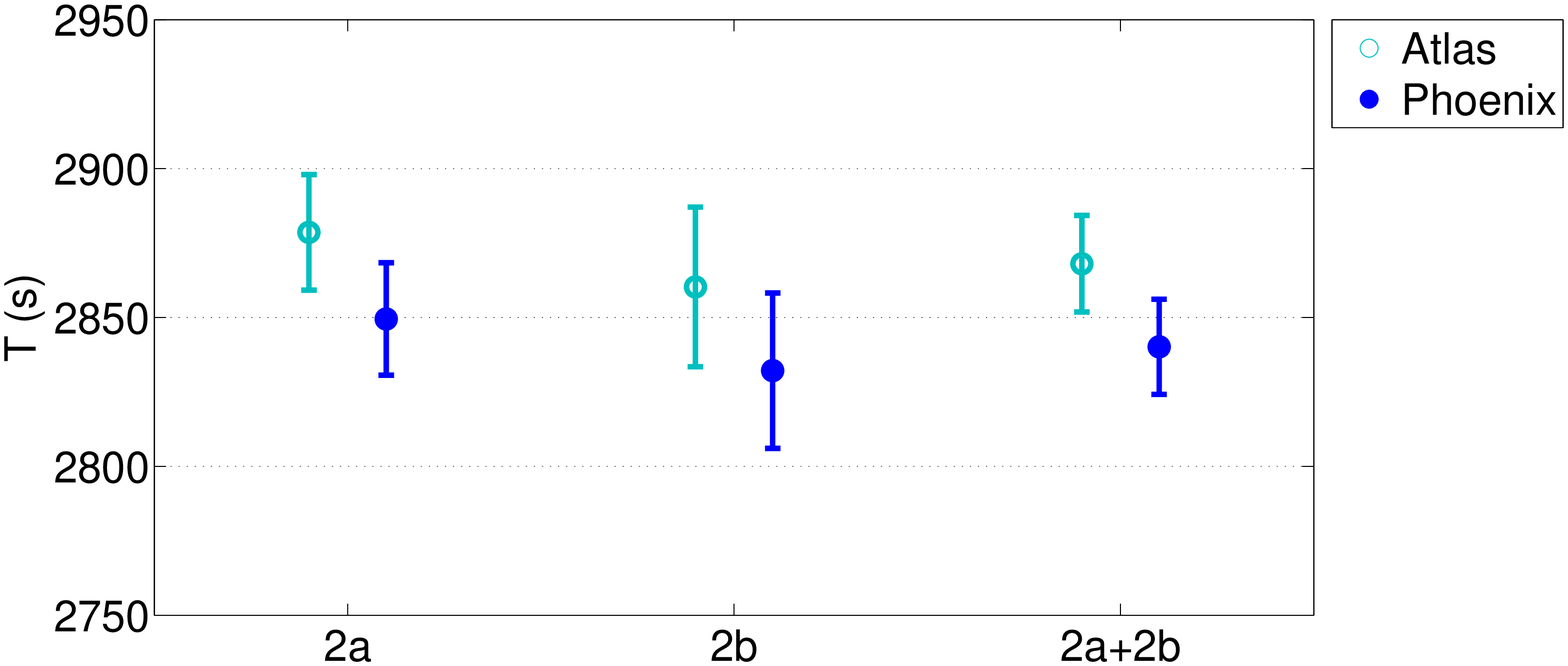}
\caption{Comparisons of the parameters $b$ and $T$, obtained in this paper, with common orbital parameters for observations at the same wavelength, and Atlas or Phoenix stellar models. \label{fig9}}
\end{figure}
\begin{figure}[!h]
%\epsscale{0.80}
\plotone{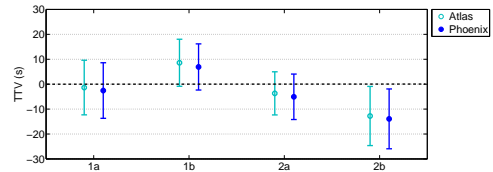}
\caption{Best fitted time-shifts of the mid-transit with respect to the periodically predicted mid-transit times, assuming Atlas or Phoenix stellar models. Note that results are model-independent, because mid-transit time will not correlate with limb darkening parameters, or any other physical parameters. \label{fig10}}
\end{figure}

\clearpage

\section{Discussion}

\subsection{Comparing observations}

\begin{figure}[!h]
%\epsscale{0.80}
\plotone{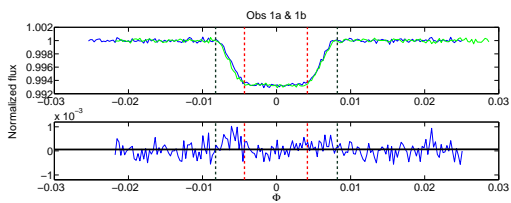}
\caption{Top panel: detrended light-curves for Obs 1a (blue), and for Obs 1b (green). Bottom panel: Residuals between the two observations. Black dotted lines delimit the out-of-transit, red dotted lines delimit the in-transit (as defined in Sec. \ref{ssec:pcc}). \label{fig11}}
\end{figure}

\begin{figure}[!h]
%\epsscale{0.80}
\plotone{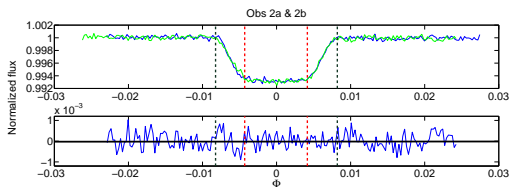}
\caption{Top panel: detrended light-curves for Obs 2a (blue), and for Obs 2b (green). Bottom panel: Residuals between the two observations. Black dotted lines delimit the out-of-transit, red dotted lines delimit the in-transit (as defined in Sec. \ref{ssec:pcc}). \label{fig12}}
\end{figure}

Fig. \ref{fig11} and \ref{fig12} report the superpositions of 3.6 and 4.5 $\mu$m light-curves respectively, and the residuals. In both cases the mean value of the in-transit residuals is small ($\lesssim$5$\times$10$^{-5}$), but the transit 1b is clearly longer than transit 1a, as measured by transit duration ($T$) parameters. As $T$ is function of the orbital parameters and stellar model, this is the reason why simultaneous fits with common orbital parameters and stellar model do not behave very well. We also note that the ingresses of transits 2a and 2b have different slopes.

%\subsection{Comparing 3.6 and 4.5$\mu$m transits}
The difference between $p^2$ values at the two wavelengths is (on average) $\sim$5$\times$10$^{-5}$, then there is no evidence of differences in planetary atmosphere's absorption at the two wavelengths.

The orbital parameters at the two wavelengths are also comparable, as detailed in Sec. \ref{sec:starting_fits} and \ref{sec:combined_fits}. Simultaneous fits over the four observations with common orbital parameters do not add any information.

\subsection{Comparison with previous analyses of the same observations}

Fig. \ref{fig7} reports the parameter values obtained in this paper for the individual observations with the analogues reported by \cite{bea11, knu11}.  Our results suggest a constant value of the transit depth (largely within 1$\sigma$) both between the 3.6 and 4.5 $\mu$m observations, and for the two wavelengths. \cite{knu11} report variations of the transit depth with a 3.4$\sigma$ significance between the two epochs at 3.6 $\mu$m, and 2.1$\sigma$ at 4.5 $\mu$m, which they attributed to stellar activity. \cite{bea11} also obtained significant differences between different epochs at the same wavelength, but they attributed such discrepancies to an unfavorable transit-systematic phasing, then they discarded those epochs from the analysis. Our error bars are generally comparable to the ones reported in both previous papers, but in some cases they are larger up to a factor $\sim$2. This is not surprising, because we are not making any prior assumptions about the signals, to guarantee a high degree of objectivity \citep{mor14, wal12, wal13}. We conclude that our detrending method lead to more robust results than the previous ones in the literature, and they show no evidence of stellar activity variations at $\sim$10$^{-4}$ photometric level. Recent results from Hubble/WFC3 observations at 1.2$-$1.6 $\mu$m \citep{knu14} also show no significant transit depth variations over four observations in about 2 months.

\subsection{Comparison with other observations}

\begin{figure}[!h]
%\epsscale{0.80}
\plotone{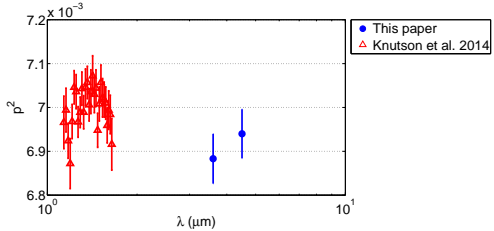}
\caption{Transit depth values obtained in this paper at 3.6 and 4.5 $\mu$m (blue circles); values at 1.2$-$1.6 $\mu$m reported by \cite{knu14} (red triangles). \label{fig13}}
\end{figure}

Fig. \ref{fig13} compares our estimated transit depth values at 3.6 and 4.5 $\mu$m (averaged over the observations at the same wavelength) with the most recent results at 1.2$-$1.6 $\mu$m \citep{knu14}. The resulting spectrum is featureless within the error bars. However, when comparing transit depth measurements at different wavelengths, we should ensure that the GJ436 system is uniformly modelled, i.e. same stellar model and orbital parameters. A uniform multiwavelength reanalysis is required to confirm this result, and investigate potential small features.

Our non-detection of TTVs higher than $\sim$30 s (see Sec. \ref{ssec:TTV}) is consistent with previous analyses in the infrared \citep{alo08, cac09, pont09, bal10, knu11, knu14}. We measured a significant TDV ($\sim$80 s) between Obs 1a and 1b; we did not find any study of TDVs for GJ436b in the literature, but injecting parameters from \cite{knu11} into our Eq. \ref{eqn:T} we obtain a similar trend for the same observations. More observations are required to investigate the cause of the apparent TDV between Obs 1a and 1b, whether it is due to a perturber (as currently required to explain the high orbital eccentricity), a stellar phenomenon, or something else.

\section{Conclusions}

We have applied a blind signal-source separation method, firstly proposed by \cite{mor14}, to analyze other photometric data of primary transits of an exoplanet, and extending its validity to the Spitzer/IRAC 4.5 $\mu$m band. These datasets were more challenging to analyze, because of the lower transit depth, comparable with the amplitude of the instrumental pixel-phase signal, the transit duration, very similar to the period of said signal, and possible stellar variability.

We obtain consistent results between transits at different epochs, ruling out stellar activity variations within $\sim$10$^{-4}$ photometric level. We do not detect any significant difference for the transit depth at 3.6 and 4.5 $\mu$m, neither with the recent measurements at 1.2$-$1.6 $\mu$m \citep{knu14}, supporting the hypothesis of a flat transmission spectrum. We measure a TDV of $~$80 s between transits separated by 7 orbits (2 $\sigma$ significance level), but no significant TTVs; more measurements are required to investigate the possible presence of a perturber, and its nature. Also, more uniform analyses at other wavelengths are required to get a more reliable transmission spectrum.\\

G. Morello is funded by UCL Perren/Impact scholarship (CJ4M/CJ0T). I. P. Waldmann is funded by the European Research Council Grant ``Exolights''. G. Tinetti is funded by the Royal Society. G. Micela is supported by ``Progetto Premiale - A Way to Other Worlds'', funded by Italian Minister for University and Scientific Research''.

\appendix

\section{ICA}
\subsection{Rationale}
\label{sec:app0added}

ICA is a special case of ``blind source separation'' technique, i.e. it aims to separate original source signals from observations with minimal assumptions. The assumptions for standard  ICA \footnote{There are some variants of ICA which do include prior information, e.g. \cite{bar11,igu02,sto02}.} are:
\begin{enumerate}
\item the source signals are statistically independent;
\item observations are linear mixtures of the source signals;
\item the number of distinguishable observations is not smaller than the number of sources.
\end{enumerate}

The first condition is easily verified if the signals have different origins, i.e. the astronomical target, other background objects, and the instruments. Additionally, some studies found that ICA algorithms can separate also signals that are not exactly independent \citep{hyv00, hyv01}.

The second condition is more questionable, given that some instrumental systematics might be multiplicative rather than additive. Alternative ICA algorithms consider non-linear mixing of the source signals, but some additional information are required to perform the separation, and, in general, there is not a unique solution \citep{hyv01}. Based on the following evidences, we found that, for Spitzer/IRAC light-curves, the classic assumption of linear mixing leads to reliable and robust results:
\begin{itemize}
\item detrended light-curves present a low level of residual scatter, compared to the literature \citep{bea11, knu11};
\item planetary and stellar parameters measured at different epochs are consistent (this is not a necessary condition);
\item non-transit components have the same characteristics, e.g. periodicity and amplitude, of known instrumental systematics.
\end{itemize}
This seems to be in contrast with the standard (empirical) pixel-phase effect method used to detrend Spitzer data \citep{faz04}: flux measurements are correlated with the position of the centroid on a pixel, the cause of this is assumed to be an intra-pixel sensitivity variation, hence the systematics model is multiplied to the astrophysical signal. We are now investigating this question through simulated observations (Morello et al., in prep.); we report here our preliminary results:
\begin{itemize}
\item either inter- and intra-pixel effects (or both) can originate systematics similar to the ones observed in Spitzer;
\item inter-pixel effects are additive, as in our ICA model;
\item intra-pixel effects are not additive, but the ICA algorithm is still able to significantly reduce their presence in the light-curves (our simulations currently indicate a reduction by a factor of 7 for the amplitudes of systematic components from an original 3.5$\times$10$^{-3}$ photometric level, outperforming the pixel-phase method by a factor of 2.3-3.3).
\end{itemize}

The third condition is case-dependent, since the number of components is not known a priori, and the number of pixels is limited by the width of the PSF. Also, if all the pixels contain the same systematic signals with the same weights relative to the astrophysical signal, the pixel-light-curves would not be distinguishable, and separation would be impossibile. Given the results obtained, we infer that we have a sufficient number of distinguishable pixel-light-curves to detrend our signals up to a 2$\times$10$^{-4}$ photometric precision.

\subsection{Performances of MULTICOMBI algorithm}

In this section, we discuss the ability of ICA to separate different kinds of signals. It depends on the particular algorithm used, in our case MULTICOMBI \citep{tic08}.  MULTICOMBI is a powerful tool, that optimally mixes two complementary algorithms, i.e. EFICA \citep{kol06}, designed to separate non-gaussian signals, and WASOBI \citep{yer00}, specialized to separate gaussian auto-regressive and time-correlated components.

\begin{figure}[!h]
\epsscale{0.80}
\plotone{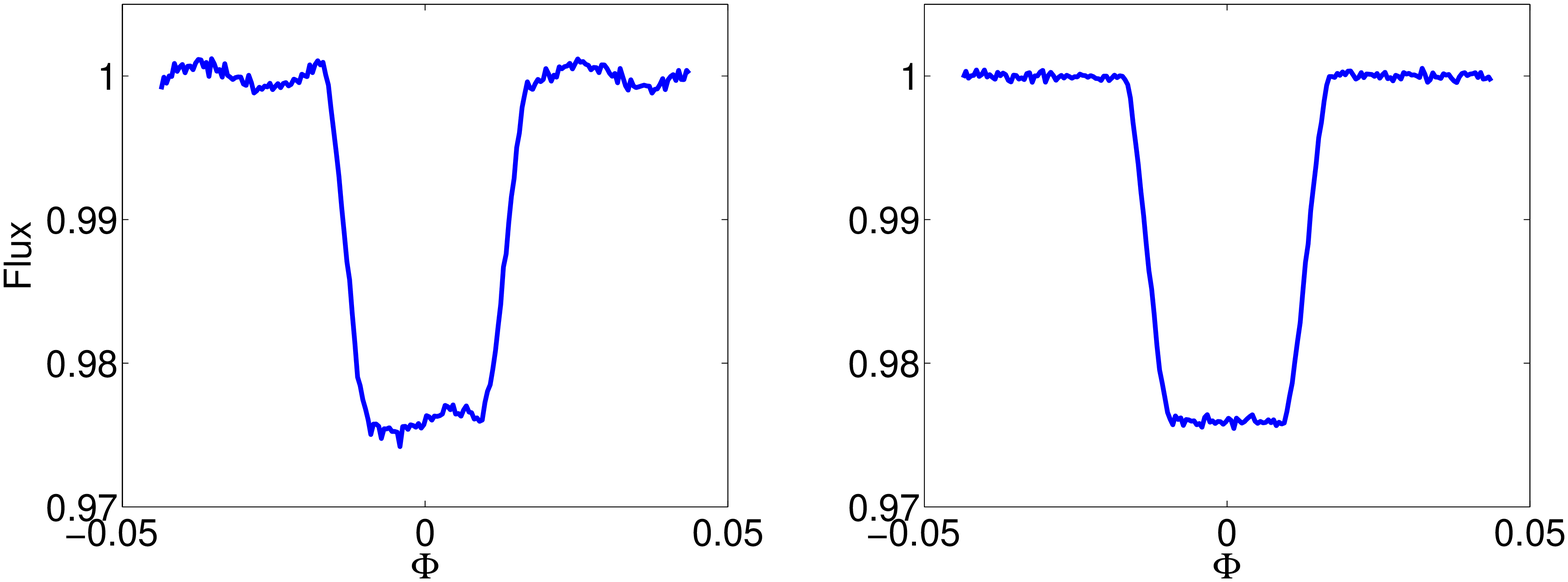}
\plotone{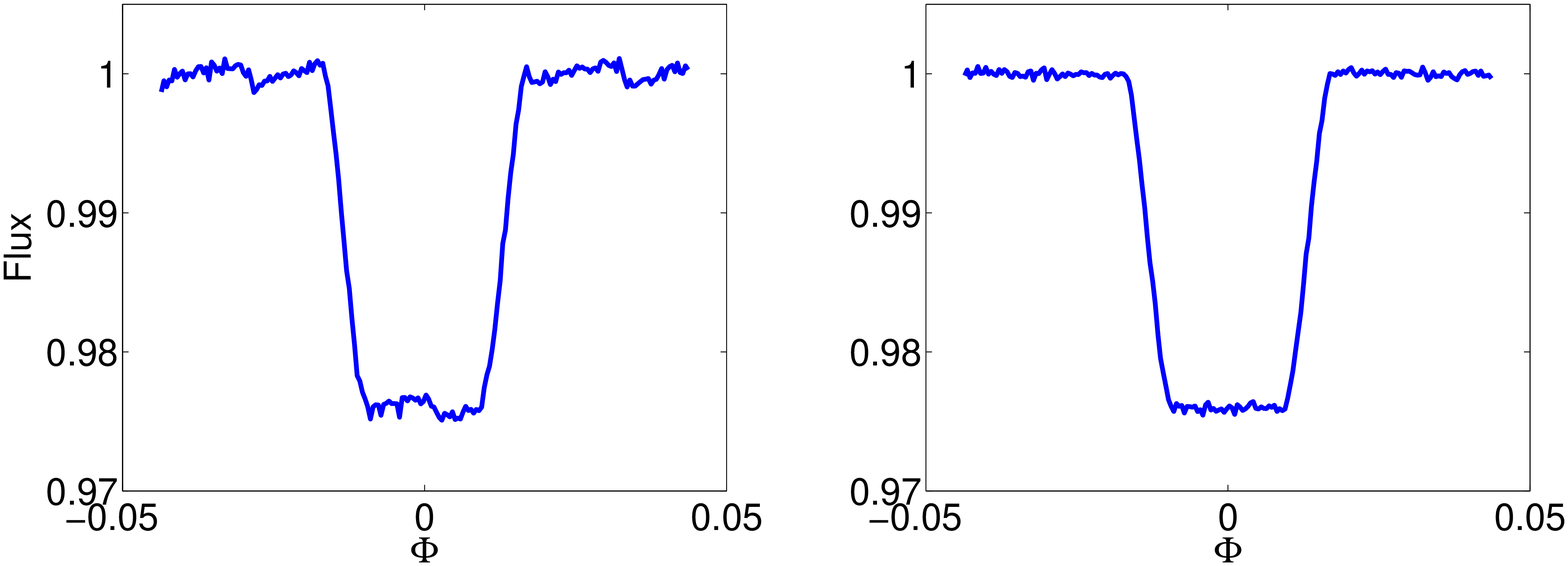}
\plotone{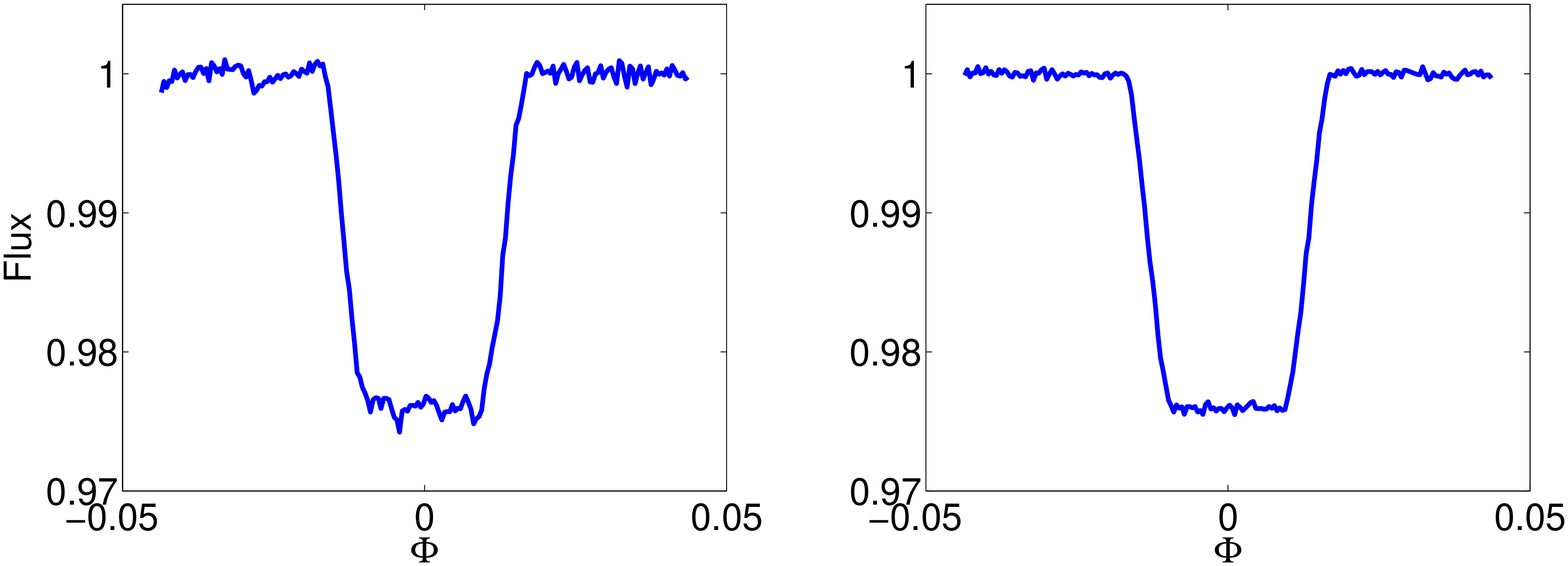}
\plotone{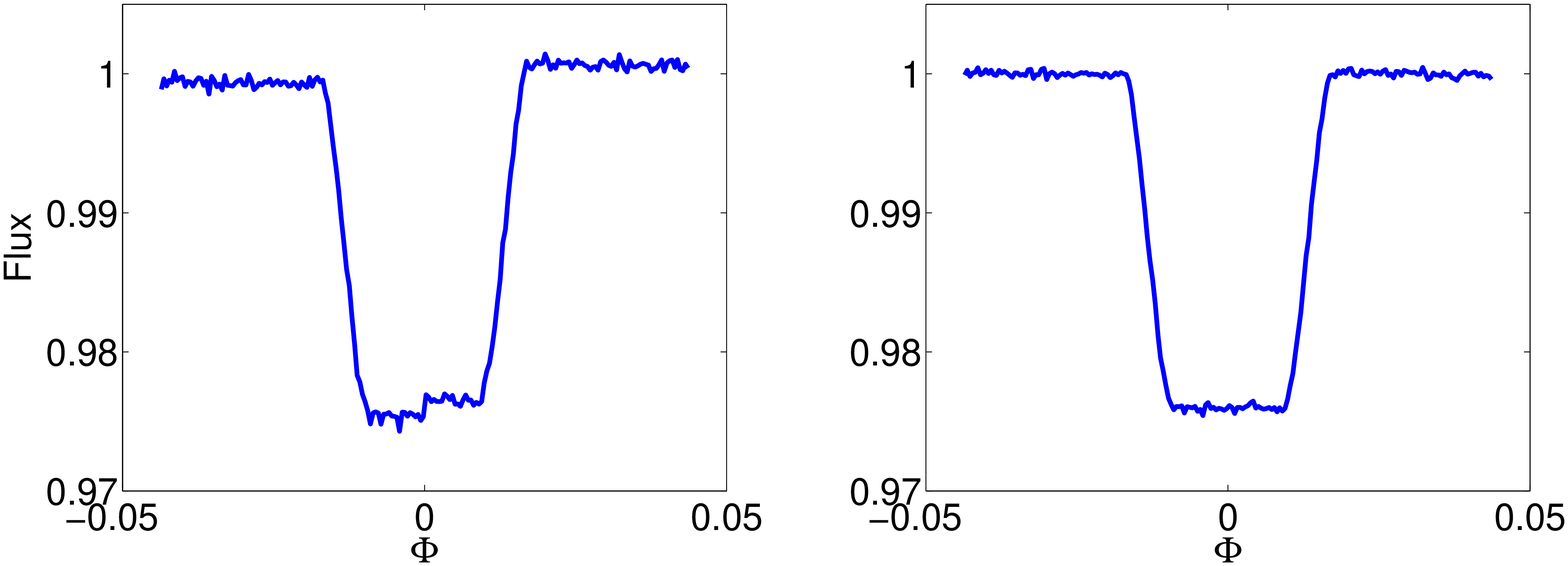}
\caption{Left panels: simulations of normalized raw ligh-curves with different systematic effects, due to pointing jitter and different pixel responses. Jitter time series are, from top to bottom: sinusoidal, periodic sawtooth (smoothed), successive smooth sawtooths with increasing frequency, and sudden shift. Right panels: Correspondent detrended light-curves with pixel-ICA method. Retrieved parameters are consistent with the values adopted to generate the raw light-curves. \label{fig14}}
\end{figure}

\begin{figure}[!h]
\epsscale{0.80}
\plotone{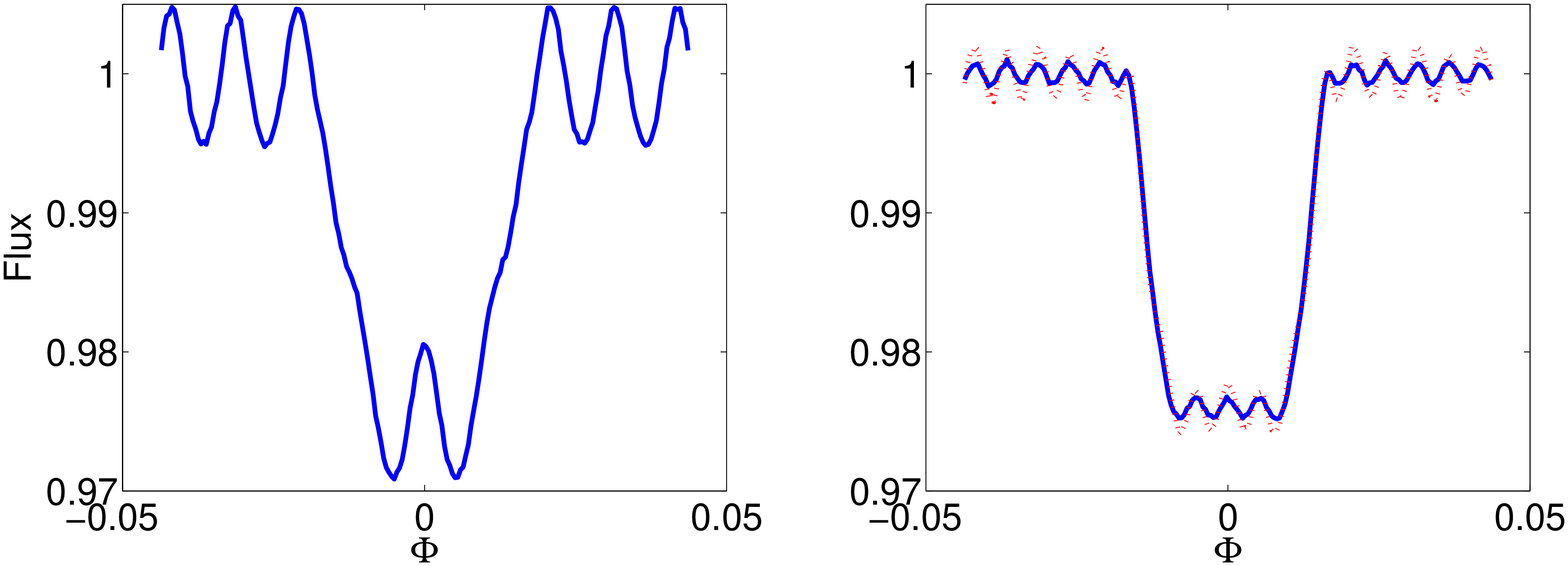}
\plotone{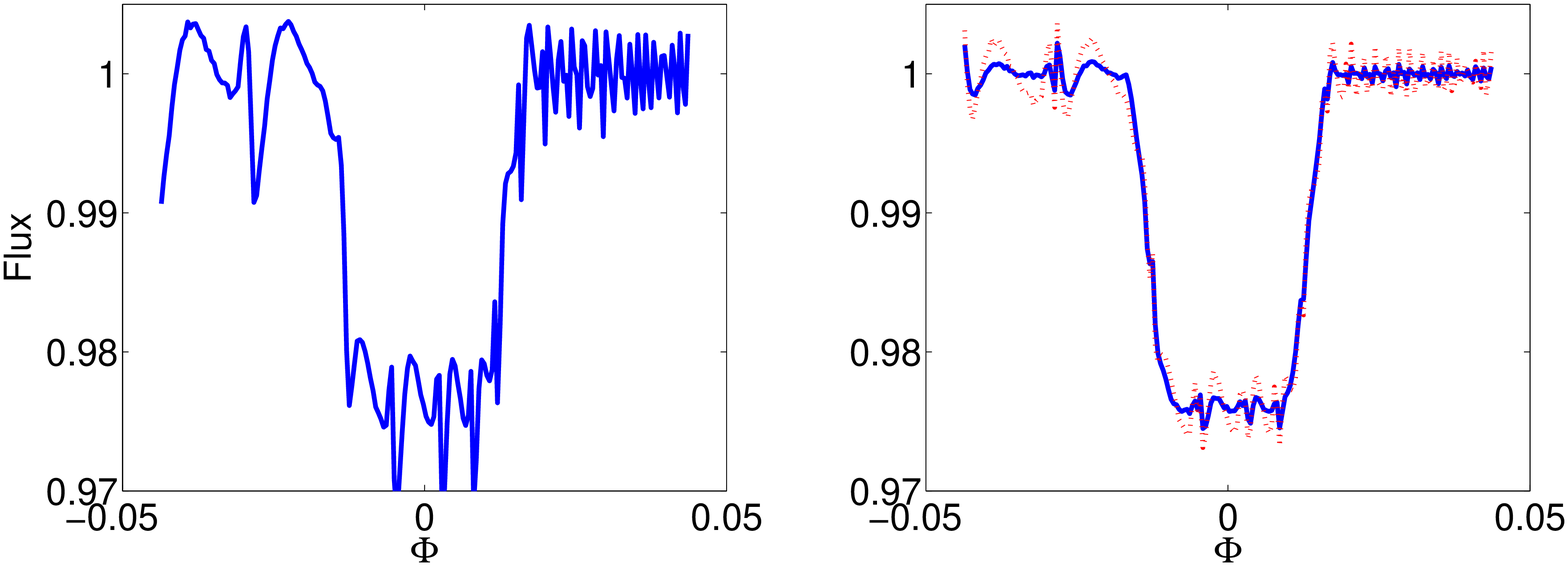}
\caption{Left panels: simulations of normalized raw ligh-curves with different systematic effects, due to pointing jitter and strong intra-pixel variations. Jitter time series are: (top) sinusoidal, (bottom) successive smooth sawtooths with increasing frequency. Right panels: Correspondent detrended light-curves with (blue) pixel-ICA method, and (red, dashed) traditional pixel-phase method. Note that ICA method outperforms the pixel-phase method: lower residual systematics and retrieved parameters closer to the true values. \label{fig15}}
\end{figure}

We tested MULTICOMBI performances with simulated observations of planetary transits affected by a large variety of systematic signals, including non-stationary signals with changing frequencies and amplitudes, sudden change points, transient behaviours, and long-term monotonic drifts. In all cases the algorithm successfully dentrended the systematic components, except monotonic drifts (see Fig. \ref{fig14} and \ref{fig15} for some examples). Spitzer/IRAC lightcurves typically start with a drift before stabilization; this would explain the improved performance if rejecting part of the earlier data points.

A more detailed analysis of the performances of the ICA detrending method adopted in this paper for different instrument cases, i.e. inter- or intra-pixel effects, amplitude and frequency of the systematics, temporal structures, non-stationarity, individual pixel peculiarities, is ongoing (Morello et al., in prep.).

The calculation time for a MULTICOMBI session is, in our cases less than 3 s. We are considering a number of signals $d=$25, and N$\sim$10$^{3}$ data points. For non-binned datasets, i.e. $N\sim$10$^{5}$ data points, the calculation time varies in the range 45-105 s, then it is, in general, case dependent. We measured the computation time for different values of $d$ and $N$, and found that the algorithm complexity scales as $\mathcal{O}(d^2N)$, as predicted for EFICA in \cite{kol06}.

\clearpage

\subsection{ICA errors}
\label{sec:app0}

If ICA were able to separate the original source signals perfectly, the parameter error bars would be fully determined by the residual scatter on the detrended light-curve. In general, we expect this not to be the case, since any detrending method would introduce some bias in the parameter estimates. We model such unknown bias as an additive uncertainty, $\sigma_{ICA}$, in the time series, leading to Eq. \ref{eqn:sigmapar}.
\cite{mor14} report the following formula for $\sigma_{ICA}$:
\begin{equation}
\label{eqn:sigmaica}
\sigma_{ICA}^2 = f^2 \left ( \sum_{j} o_j^2 \textbf{ISR}_j + \sigma_{ntc-fit}^2 \right )
\end{equation}
where $\textbf{ISR}$ is the so-called Interference-to-Signal-Ratio matrix, $o_j$ are the coefficients of the non-transit-components, $m$ is their number, $\sigma_{ntc-fit}$ is the standard deviation of residuals from the referent raw light-curve, out of the transit, $f$ is the normalising factor for the detrended light-curve. The sum on the left takes into account the precision of the components extracted by the algorithm; $\sigma_{ntc-fit}$ indicates how well the linear combination of components approximates the out-of-transit. Note that, while the first term increases with the number of components considered (see Sec. \ref{ssec:errors}), the second term decreases. The optimal strategy is to remove all the extracted non-transit components from the raw light-curve, though many results obtained by removing the most significant components (to be determined) are almost identical (see \cite{mor14}, Sec. 2.5.2).

MULTICOMBI code produces two Interference-to-Signal-Ratio matrices, $\textbf{ISR}^{EF}$, associated to the algorithm EFICA, and $\textbf{ISR}^{WA}$, associated to the algorithm WASOBI.
In \cite{mor14} we estimated the global $\textbf{ISR}$ as the arithmetic mean of $\textbf{ISR}^{EF}$ and $\textbf{ISR}^{WA}$; this a very conservative estimate, which does not take into account the outperforming separation capabilities of MULTICOMBI compared to EFICA and WASOBI; here we suggest a more appropriate definition:
\begin{equation}
\label{eqn:ISR}
\textbf{ISR}_{i,j} = min \left ( \textbf{ISR}_{i,j}^{EF}, \textbf{ISR}_{i,j}^{WA} \right )
\end{equation}

In the cases analyzed in \cite{mor14} the contributions of the $\textbf{ISR}$ terms were $\sim$10$\%$ of the total error bars, then adopting the new definition of $\textbf{ISR}$ would not modify the results significantly. Here, we find that the $\textbf{ISR}$ contributions to the error bars are comparable with the other terms, probably because some instrumental systematics and the transit signals have similar timescales and amplitudes, making the separation more uncertain. Tab. \ref{tab3} reports the values of $\sigma_{ICA}$ obtained for each observations, with $\textbf{ISR}$ calculated according to Eq. \ref{eqn:ISR}, and according to the arithmetic mean definition.

\begin{table}[!h]
\begin{center}
\caption{Estimated $\sigma_{ICA}$ values for the four observations (Eq. \ref{eqn:sigmaica} and \ref{eqn:ISR}), and worst case values according to a more conservative estimate of $\textbf{ISR}$ (arithmetic mean of $\textbf{ISR}^{EF}$ and $\textbf{ISR}^{WA}$). \label{tab3}}
\begin{tabular}{ccccc}
\tableline\tableline
Obs. number & $\sigma_{ICA}$ & $\sigma_{ICA}$ (max)\\
\tableline
1a & 4.24$\times$10$^{-4}$ & 5.97$\times$10$^{-4}$\\
1b & 3.41$\times$10$^{-4}$ & 4.52$\times$10$^{-4}$\\
2a & 2.79$\times$10$^{-4}$ & 3.07$\times$10$^{-4}$\\
2b & 3.97$\times$10$^{-4}$ & 6.42$\times$10$^{-4}$\\
\tableline
\end{tabular}
%% Any table notes must follow the \end{tabular} command.
\end{center}
\end{table}

The error bars obtained in this paper with the definition in Eq. \ref{eqn:ISR} are consistent with results from the same tests of robustness reported in \cite{mor14}, i.e. different number of components, pixel arrays, and partial time series. For completeness, tables in App. \ref{tables} report the error bars obtained from residual scatter only, including $\sigma_{ICA}$ with \textbf{ISR} defined as in Eq. \ref{eqn:ISR}, and with the previous definition of \textbf{ISR}. The relative difference between error bars obtained with the two definitions of \textbf{ISR} is 0$-$30$\%$, which, in general, may be important for atmospheric characterization, but it does not affect the conclusions obtained in this paper.

\section{Full datasets analysis}
\label{sec:app1}

\begin{figure}[!h]
\epsscale{0.70}
\plotone{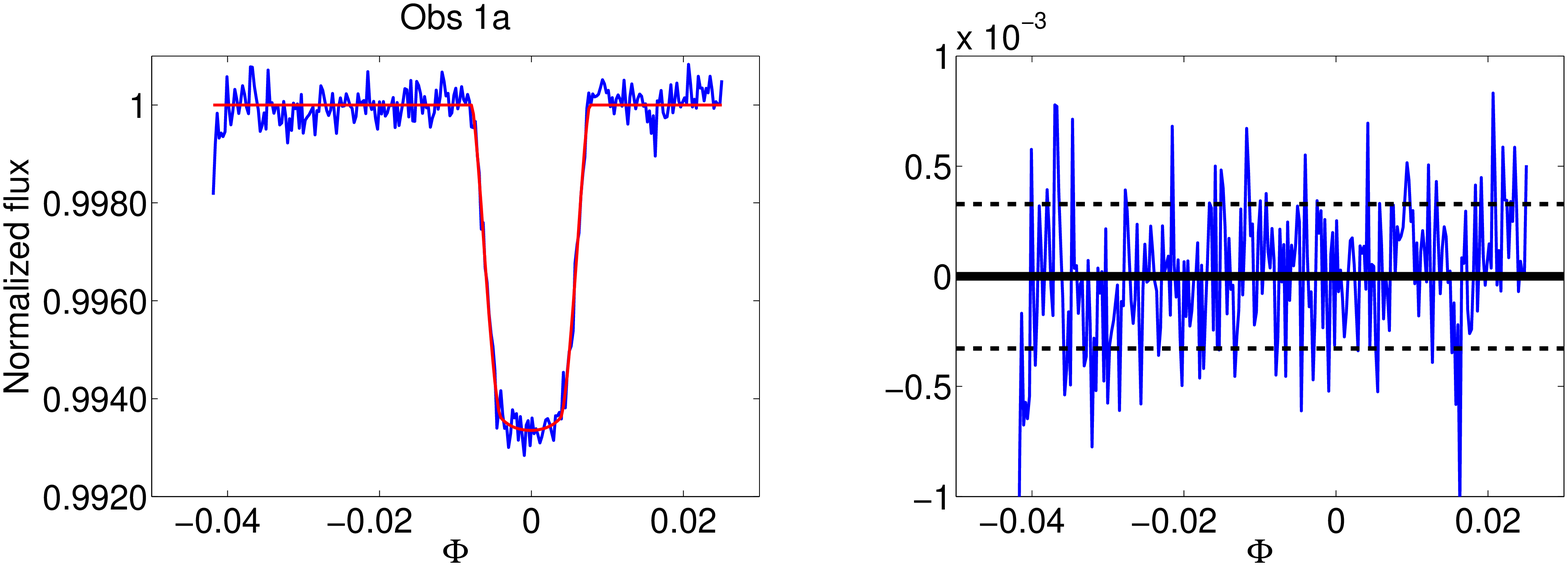}
\plotone{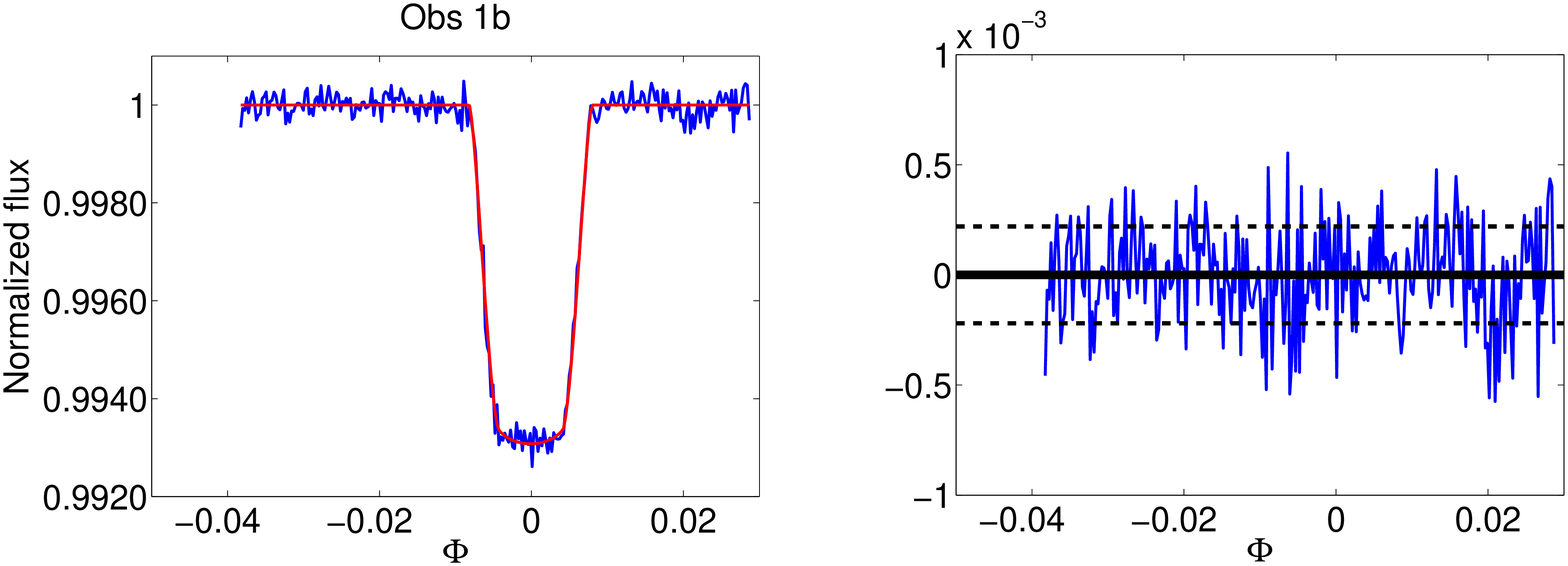}
\plotone{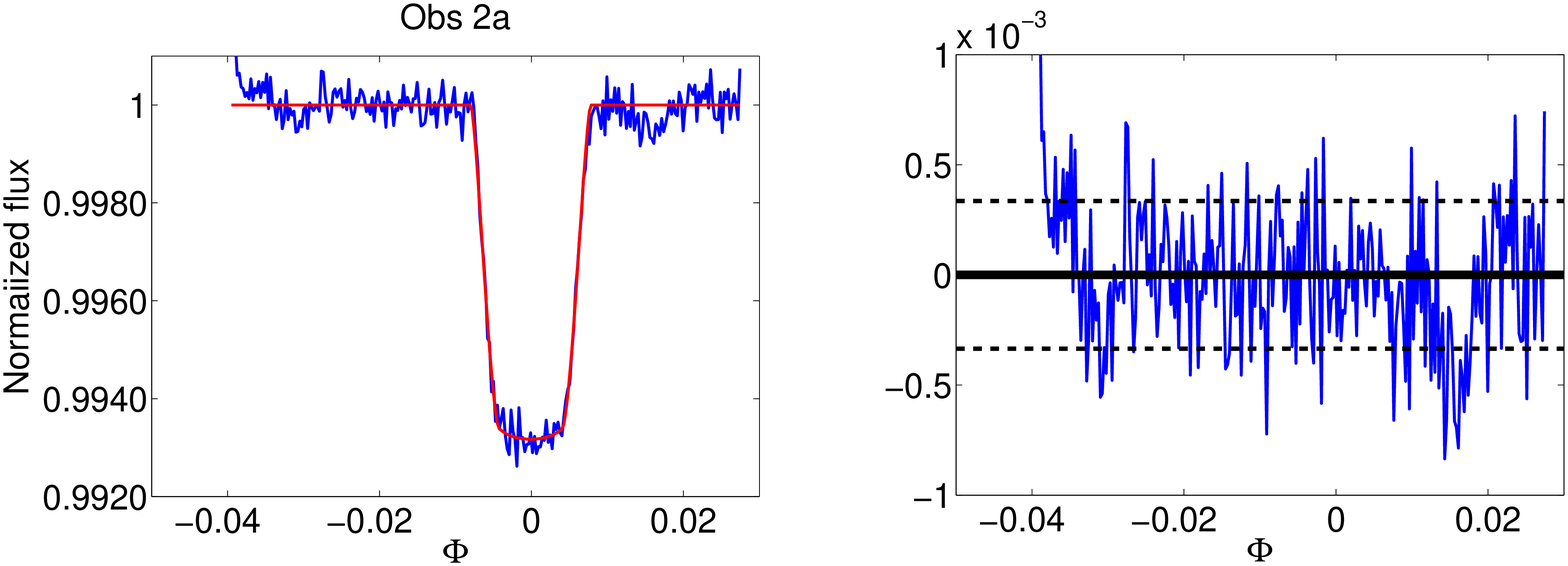}
\plotone{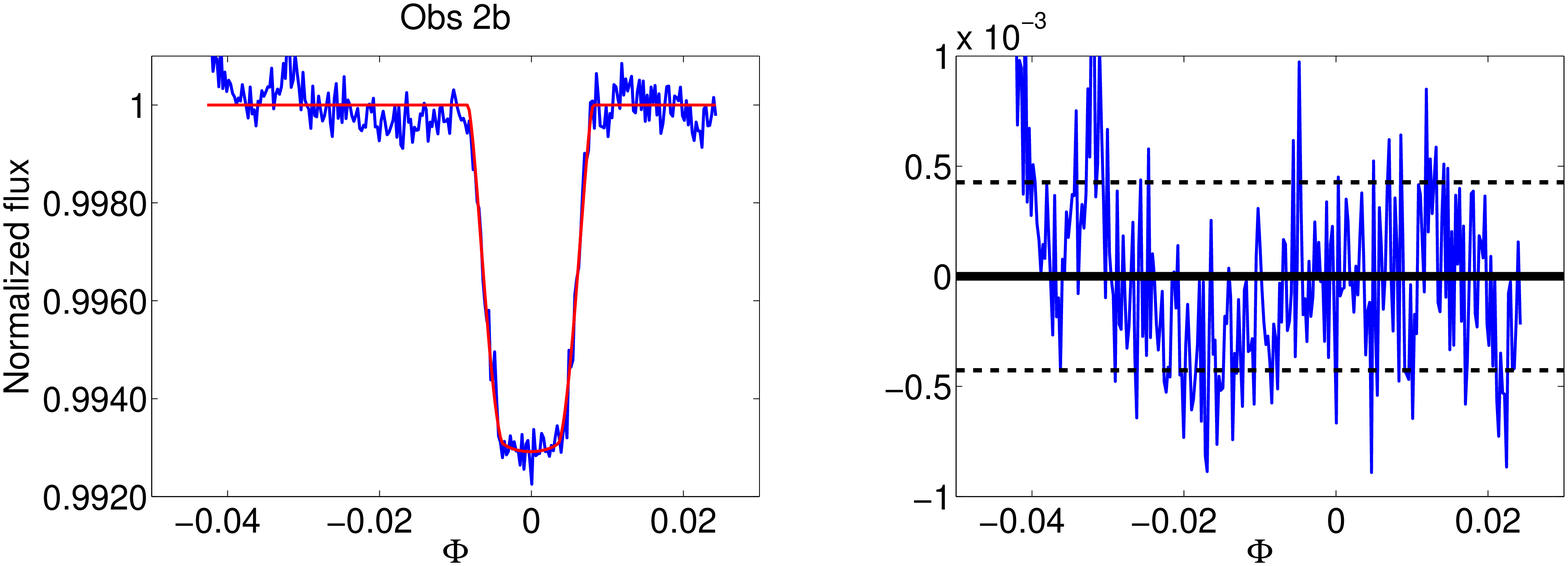}
\caption{Left panels: (blue) detrended light-curves for the four observations without data rejection, (red) best transit models overplotted, binned over 7 points; best transit models are calculated with $p$, $a_0$, and $i$ as free parameters, and Phoenix quadratic limb darkening coefficients (see Sec. \ref{ssec:ldc}). Right panels: Residuals between detrended light-curves and best transit models; black horizontal dashed lines indicate the standard deviations of residuals. \label{fig16}}
\end{figure}

Fig. \ref{fig16} reports the detrended light-curves obtained using the whole datasets. As stated in Sec. \ref{ssec:application}, the systematics originally present in the raw light-curves are greatly reduced (see Fig. \ref{fig1}), but there are still visible trends at all phases of the transits. Quantitative measurements of these trends, based on the correlations with the pixel-phase, have been discussed in Sec. \ref{ssec:pcc}, together with the trends obtained by rejecting the first 450 data points before processing. Also, standard deviation of residuals between the light-curves and the transit models are larger for the cases with no preliminary data rejection (but for Obs 1b the difference is not significant). Removing the first 450 data points after ICA processing may reduce the scatter, but it does not improve the reliability of the light-curves, because trends are present at all times.

Results extracted from these worse-quality light-curves are less robust, but they are consistent with accepted results within 1 $\sigma$. Also, error bars are largely underestimated if neglecting the uncertainties due to the detrending process. It may induce to erroneously detect inter-epoch transit depth variations.

\section{Alternative transit model-fits}
\label{sec:app2}

\subsection{Free limb darkening}

\begin{figure}[!h]
\epsscale{0.80}
\plotone{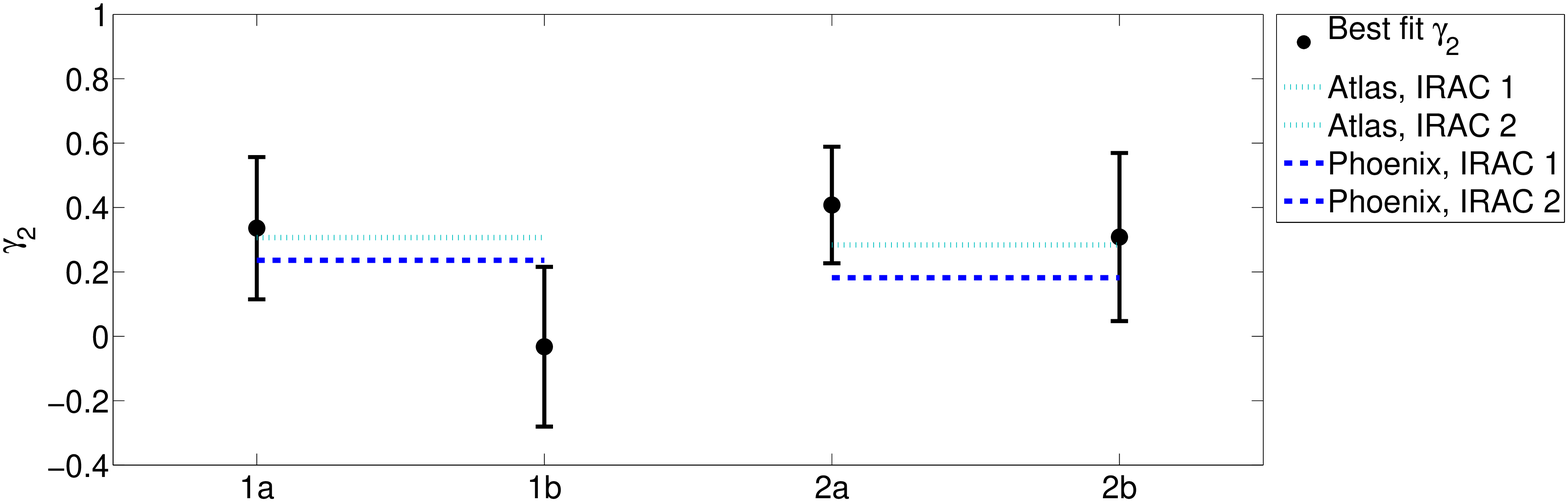}
\caption{Best fitted values of the quadratic limb darkening coefficient ($\gamma_2$) for the four observations, assuming $\gamma_1 = 0$. \label{fig17}}
\end{figure}

We performed transit model-fits with one free limb darkening parameter (linear or quadratic) in addition to the other free parameters ($p$, $a_0$, and $i$). The standard deviations of residuals between the detrended light-curves and the transit models do not change, among the models obtained with Atlas, Phoenix, free linear and free quadratic limb darkening coefficients. The parameter error bars are larger by factors in the range 1$-$3 for the free quadratic case, even much larger for the free linear case. Best parameter estimates may be more affected by intercorrelations. The pure quadratic limb darkening is a better approximation of the real case, because in both Atlas and Phoenix models the quadratic coefficients are greater than the linear ones. Fig. \ref{fig17} reports the estimates for the quadratic limb darkening coefficients, $\gamma_2$.

Note that:
\begin{enumerate}
\item they are comparable (within 1$\sigma$) with the theoretical values;
\item error bars are larger than the differences between Atlas and Phoenix values;
\item error bars do not allow to distinguish the values at 3.6 and 4.5 $\mu$m.
\end{enumerate}

Interestingly, the best estimate of $\gamma_2$ for Obs 1b is the most distant from the other values (but within 1$\sigma$), and the correspondent transit duration is now equal for Obs 1a and 1b. Although the value $\gamma_2 \simeq$0 is not reliable, it is important to note that the measured transit duration, as defined in Eq. \ref{eqn:T}, depends on the stellar intensity distribution and limb darkening model adopted, then:
\begin{itemize}
\item stellar variability is a possible cause for observed TDVs;
\item TDVs measured from observations at different wavelengths must be taken carefully.
\end{itemize}

\subsection{Free eccentricity}

We performed transit model-fits with free eccentricity, $e$, in addition to the other free parameters ($p$, $a_0$, and $i$). Eccentricity has not a great impact on the transit models: best estimates of the other parameters do not change significantly with respect to previous ones (with $e=0.16$), although best estimates for the eccentricity for different observations varies over a large range (0.08$-$0.22). Also, residuals between light-curves and models are not affected.

\subsection{Free argument of periastron}

We performed transit model-fits with free argument of periastron, $\omega$, in addition to the other free parameters ($p$, $a_0$, and $i$). They do not constrain $\omega$ very well ($\sigma_{ \omega} \sim$24$-$34); $a_0$ and $i$ are strictly correlated with $\omega$, and their error bars are $\sim$3 times larger than ones obtained with $\omega$ fixed. Also, the distributions of $\omega$, $a_0$, and $i$, are asymmetric, because best $\omega$ values are close to the edge of the range of admissible values. It is interesting to note that other parameters, such as $p$, $b$, and $T$, are not affected by $\omega$ degeneracies: their posterior distributions are indistinguishable from the ones obtained with $\omega$ fixed.

\subsection{Free phase-shift}

We performed transit model-fits with a free phase-shift in addition to other free parameters, in order to investigate the effect of possible timing variations. Results are very similar for each fitting configuration, with no evidence of timing variation below $\sim$30 s, as stated in Sec. \ref{ssec:TTV}. The free phase-shift does not affect other parameter estimates, except in the cases with free argument of periastron ($\omega$): timing variations and orbit precession are highly correlated.

\subsection{Fitting $p$, $a_0$, and $i$ with components' coefficients}

We experimented an alternative method to estimate the transit parameters and the coefficients of the independent components simultaneously, by modeling the raw lightcurves as linear combinations of the components plus a transit model. In this way, we can investigate possible correlations between transit parameters and mixing coefficients, and test the stability of an ICA model over the whole observation. If results were significantly different than the ones obtained by estimating the mixing coefficients on the out-of-transit only, it would indicate that something different has happened during the transit, either astrophysical or instrumental in nature. For the datasets analyzed in this paper, results are consistent with accepted values within 1 $\sigma$. It is worth to note that partial error bars, $\sigma_{par,0}$, obtained from the MCMCs are similar to the ones obtained with $p$, $a_0$, and $i$ only free parameters (in some cases even smaller). This indicates that the main cause of uncertainty attributed to the detrending method is not given by the mixing coefficients, but the intrinsic errors on the components extracted.

\section{Tables}
\label{tables}

\begin{table}
\begin{center}
\caption{Transit parameter estimates for the four observations, by fitting $p$, $a_0$, and $i$ as free parameters, with ATLAS quadratic limb darkening coefficients (Tab. \ref{tab2}). We report the partial error bars obtained by the residuals, the final error bars, and the worst case error bars (Eq. \ref{eqn:sigma_par}, Sec.  \ref{ssec:errors} and App. \ref{sec:app0}). \label{tab4}}
\resizebox{0.8 \textwidth}{!}{%
\begin{tabular}{cccccc}
\tableline\tableline
Obs. number & Parameters & Best values & 1-$\sigma$ errors & 1-$\sigma$ errors & 1-$\sigma$ errors\\
 & & & (residual scatter only) & (ICA) & (ICA worst case)\\
\tableline
 & $p$ & 0.0834 & 3$\times$10$^{-4}$ & 7$\times$10$^{-4}$ & 9$\times$10$^{-4}$\\
 & $a_0$ & 14.24 & 0.20 & 0.46 & 0.61\\
 & $i$ & 86.63 & 0.07 & 0.16 & 0.21\\
1a & $p^2$ & 0.00696 & 5$\times$10$^{-5}$ & 1.1$\times$10$^{-4}$ & 1.5$\times$10$^{-4}$\\
 & $b$ & 0.836 & 0.005 & 0.012 & 0.016\\
 & $T$(s) & 2835 & 10 & 24 & 33\\
 & $\sigma_0$ & 2.01$\times$10$^{-4}$ & 0.10$\times$10$^{-4}$ & & \\
\tableline
 & $p$ & 0.0836 & 3$\times$10$^{-4}$ & 6$\times$10$^{-4}$ & 7$\times$10$^{-4}$\\
 & $a_0$ & 13.93 & 0.19 & 0.39 & 0.48\\
 & $i$ & 86.57 & 0.07 & 0.14 & 0.17\\
1b & $p^2$ & 0.00699 & 5$\times$10$^{-5}$ & 9$\times$10$^{-5}$ & 1.2$\times$10$^{-4}$\\
 & $b$ & 0.834 & 0.005 & 0.011 & 0.013\\
 & $T$(s) & 2916 & 10 & 20 & 25\\
 & $\sigma_0$ & 1.96$\times$10$^{-4}$ & 0.10$\times$10$^{-4}$ & & \\
\tableline
 & $p$ & 0.0833 & 3$\times$10$^{-4}$ & 5$\times$10$^{-4}$ & 5$\times$10$^{-4}$\\
 & $a_0$ & 14.24 & 0.25 & 0.37 & 0.39\\
 & $i$ & 86.66 & 0.09 & 0.13 & 0.14\\
2a & $p^2$ & 0.00694 & 6$\times$10$^{-5}$ & 8$\times$10$^{-5}$ & 9$\times$10$^{-5}$\\
 & $b$ & 0.831 & 0.007 & 0.010 & 0.011\\
 & $T$(s) & 2879 & 13 & 19 & 21\\
 & $\sigma_0$ & 2.56$\times$10$^{-4}$ & 0.13$\times$10$^{-4}$ & & \\
\tableline
 & $p$ & 0.0842 & 4$\times$10$^{-4}$ & 6$\times$10$^{-4}$ & 9$\times$10$^{-4}$\\
 & $a_0$ & 13.38 & 0.23 & 0.40 & 0.57\\
 & $i$ & 86.34 & 0.09 & 0.15 & 0.21\\
2b & $p^2$ & 0.00709 & 6$\times$10$^{-5}$ & 1.1$\times$10$^{-4}$ & 1.6$\times$10$^{-4}$\\
 & $b$ & 0.854 & 0.006 & 0.010 & 0.014\\
 & $T$(s) & 2860 & 16 & 27 & 39\\
 & $\sigma_0$ & 2.87$\times$10$^{-4}$ & 0.15$\times$10$^{-4}$ & & \\
\tableline
\end{tabular}}
%% Any table notes must follow the \end{tabular} command.
\end{center}
\end{table}

\begin{table}
\begin{center}
\caption{Transit parameter estimates for the four observations, by fitting $p$, $a_0$, and $i$ as free parameters, with PHOENIX quadratic limb darkening coefficients (Tab. \ref{tab2}). We report the partial error bars obtained by the residuals, the final error bars, and the worst case error bars (Eq. \ref{eqn:sigma_par}, Sec.  \ref{ssec:errors} and App. \ref{sec:app0}). \label{tab5}}
\resizebox{0.8 \textwidth}{!}{%
\begin{tabular}{cccccc}
\tableline\tableline
Obs. number & Parameters & Best values & 1-$\sigma$ errors & 1-$\sigma$ errors & 1-$\sigma$ errors\\
 & & & (residual scatter only) & (ICA) & (ICA worst case)\\
\tableline
 & $p$ & 0.0828 & 3$\times$10$^{-4}$ & 6$\times$10$^{-4}$ & 8$\times$10$^{-4}$\\
 & $a_0$ & 14.02 & 0.19 & 0.44 & 0.59\\
 & $i$ & 86.54 & 0.07 & 0.15 & 0.21\\
1a & $p^2$ & 0.00686 & 4$\times$10$^{-5}$ & 1.0$\times$10$^{-4}$ & 1.4$\times$10$^{-4}$\\
 & $b$ & 0.845 & 0.005 & 0.011 & 0.015\\
 & $T$(s) & 2805 & 10 & 24 & 33\\
 & $\sigma_0$ & 2.01$\times$10$^{-4}$ & 0.10$\times$10$^{-4}$ & & \\
\tableline
 & $p$ & 0.0831 & 2$\times$10$^{-4}$ & 5$\times$10$^{-4}$ & 6$\times$10$^{-4}$\\
 & $a_0$ & 13.70 & 0.16 & 0.32 & 0.41\\
 & $i$ & 86.47 & 0.06 & 0.12 & 0.15\\
1b & $p^2$ & 0.00690 & 4$\times$10$^{-5}$ & 8$\times$10$^{-5}$ & 1.0$\times$10$^{-4}$\\
 & $b$ & 0.844 & 0.004 & 0.008 & 0.010\\
 & $T$(s) & 2884 & 10 & 20 & 25\\
 & $\sigma_0$ & 1.92$\times$10$^{-4}$ & 0.10$\times$10$^{-4}$ & & \\
\tableline
 & $p$ & 0.0830 & 3$\times$10$^{-4}$ & 5$\times$10$^{-4}$ & 5$\times$10$^{-4}$\\
 & $a_0$ & 13.99 & 0.24 & 0.35 & 0.37\\
 & $i$ & 86.55 & 0.08 & 0.12 & 0.13\\
2a & $p^2$ & 0.00688 & 5$\times$10$^{-5}$ & 8$\times$10$^{-5}$ & 8$\times$10$^{-5}$\\
 & $b$ & 0.841 & 0.006 & 0.009 & 0.010\\
 & $T$(s) & 2850 & 13 & 19 & 20\\
 & $\sigma_0$ & 2.58$\times$10$^{-4}$ & 0.13$\times$10$^{-4}$ & & \\
\tableline
 & $p$ & 0.0836 & 4$\times$10$^{-4}$ & 6$\times$10$^{-4}$ & 9$\times$10$^{-4}$\\
 & $a_0$ & 13.24 & 0.22 & 0.37 & 0.54\\
 & $i$ & 86.27 & 0.08 & 0.14 & 0.20\\
2b & $p^2$ & 0.00699 & 6$\times$10$^{-5}$ & 1.0$\times$10$^{-4}$ & 1.4$\times$10$^{-4}$\\
 & $b$ & 0.861 & 0.005 & 0.009 & 0.013\\
 & $T$(s) & 2832 & 15 & 26 & 37\\
 & $\sigma_0$ & 2.88$\times$10$^{-4}$ & 0.15$\times$10$^{-4}$ & & \\
\tableline
\end{tabular}}
%% Any table notes must follow the \end{tabular} command.
\end{center}
\end{table}

\begin{table}
\begin{center}
\caption{Transit parameter estimates for the four observations, by fitting $p$ as a free parameter, $a_0$ and $i$ identical for the observations at the same wavelength, with ATLAS quadratic limb darkening coefficients (Tab. \ref{tab2}). We report the partial error bars obtained by the residuals, the final error bars, and the worst case error bars (Eq. \ref{eqn:sigma_par}, Sec.  \ref{ssec:errors} and App. \ref{sec:app0}). \label{tab6}}
\resizebox{\textwidth}{!}{%
\begin{tabular}{cccccc}
\tableline\tableline
Obs. number & Parameters & Best values & 1-$\sigma$ errors & 1-$\sigma$ errors & 1-$\sigma$ errors\\
 & & & (residual scatter only) & (ICA) & (ICA worst case)\\
\tableline
 & $p_{1a}$ & 0.0831 & 2.5$\times$10$^{-4}$ & 5$\times$10$^{-4}$ & 7$\times$10$^{-4}$\\
 & $p_{1b}$ & 0.0840 & 2.5$\times$10$^{-4}$ &  5$\times$10$^{-4}$ & 7$\times$10$^{-4}$\\
 & $a_0$ & 14.04 & 0.13 & 0.28 & 0.37\\
 & $i$ & 86.58 & 0.05 & 0.10 & 0.13\\
1a + 1b & $p_{1a}^2$ & 0.00690 & 4$\times$10$^{-5}$ & 9$\times$10$^{-5}$  & 1.1$\times$10$^{-4}$\\
 & $p_{1b}^2$ & 0.00706 & 4$\times$10$^{-5}$ & 9$\times$10$^{-5}$ & 1.2$\times$10$^{-4}$\\
 & $b$ & 0.836 & 0.004 & 0.007 & 0.010\\
 & $T$(s) & 2876 & 8 & 16 & 21\\
 & $\sigma_0$ & 2.05$\times$10$^{-4}$ & 0.07$\times$10$^{-4}$ & & \\
\tableline
 & $p_{2a}$ & 0.0838 & 3$\times$10$^{-4}$ & 5$\times$10$^{-4}$ & 6$\times$10$^{-4}$\\
 & $p_{2b}$ & 0.0838 & 3$\times$10$^{-4}$ & 5$\times$10$^{-4}$ & 6$\times$10$^{-4}$\\
 & $a_0$ & 13.74 & 0.17 & 0.26 & 0.33\\
 & $i$ & 86.47 & 0.06 & 0.10 & 0.12\\
2a + 2b & $p_{2a}^2$ & 0.00702 & 5$\times$10$^{-5}$ & 8$\times$10$^{-5}$ & 1.1$\times$10$^{-4}$\\
 & $p_{2b}^2$ & 0.00702 & 5$\times$10$^{-5}$ & 8$\times$10$^{-5}$ & 1.1$\times$10$^{-4}$\\
 & $b$ & 0.845 & 0.004 & 0.007 & 0.009\\
 & $T$(s) & 2868 & 10 & 16 & 20\\
 & $\sigma_0$ & 2.73$\times$10$^{-4}$ & 0.10$\times$10$^{-4}$ & & \\
\tableline
\end{tabular}}
%% Any table notes must follow the \end{tabular} command.
\end{center}
\end{table}

\begin{table}
\begin{center}
\caption{Transit parameter estimates for the four observations, by fitting $p$ as a free parameter, $a_0$ and $i$ identical for the observations at the same wavelength, with PHOENIX quadratic limb darkening coefficients (Tab. \ref{tab2}). We report the partial error bars obtained by the residuals, the final error bars, and the worst case error bars (Eq. \ref{eqn:sigma_par}, Sec.  \ref{ssec:errors} and App. \ref{sec:app0}). \label{tab7}}
\resizebox{\textwidth}{!}{%
\begin{tabular}{cccccc}
\tableline\tableline
Obs. number & Parameters & Best values & 1-$\sigma$ errors & 1-$\sigma$ errors & 1-$\sigma$ errors\\
 & & & (residual scatter only) & (ICA) & (ICA worst case)\\
\tableline
 & $p_{1a}$ & 0.0825 & 2$\times$10$^{-4}$ & 5$\times$10$^{-4}$ & 6$\times$10$^{-4}$\\
 & $p_{1b}$ & 0.0834 & 2$\times$10$^{-4}$ & 5$\times$10$^{-4}$ & 6$\times$10$^{-4}$\\
 & $a_0$ & 13.82 & 0.12 & 0.26 & 0.34\\
 & $i$ & 86.49 & 0.04 & 0.09 & 0.12\\
1a + 1b & $p_{1a}^2$ & 0.00681 & 4$\times$10$^{-5}$ & 8$\times$10$^{-5}$ & 1.0$\times$10$^{-4}$\\
 & $p_{1b}^2$ & 0.00696 & 4$\times$10$^{-5}$ & 8$\times$10$^{-5}$ & 1.0$\times$10$^{-4}$\\
 & $b$ & 0.845 & 0.003 & 0.007 & 0.008\\
 & $T$(s) & 2845 & 7 & 16 & 21\\
 & $\sigma_0$ & 2.04$\times$10$^{-4}$ & 0.07$\times$10$^{-4}$ & & \\
\tableline
 & $p_{2a}$ & 0.0833 & 3$\times$10$^{-4}$ & 5$\times$10$^{-4}$ & 6$\times$10$^{-4}$\\
 & $p_{2b}$ & 0.0833 & 3$\times$10$^{-4}$ & 5$\times$10$^{-4}$ & 6$\times$10$^{-4}$\\
 & $a_0$ & 13.57 & 0.16 & 0.25 & 0.31\\
 & $i$ & 86.40 & 0.06 & 0.09 & 0.11\\
2a + 2b & $p_{2a}^2$ & 0.00694 & 5$\times$10$^{-5}$ & 8$\times$10$^{-5}$ & 1.0$\times$10$^{-4}$\\
 & $p_{2b}^2$ & 0.00694 & 5$\times$10$^{-5}$ & 8$\times$10$^{-5}$ & 1.0$\times$10$^{-4}$\\
 & $b$ & 0.852 & 0.004 & 0.006 & 0.008\\
 & $T$(s) & 2840 & 10 & 16 & 20\\
 & $\sigma_0$ & 2.74$\times$10$^{-4}$ & 0.10$\times$10$^{-4}$ & & \\
\tableline
\end{tabular}}
%% Any table notes must follow the \end{tabular} command.
\end{center}
\end{table}

%% If you wish to include an acknowledgments section in your paper,
%% separate it off from the body of the text using the \acknowledgments
%% command.

%% Included in this acknowledgments section are examples of the
%% AASTeX hypertext markup commands. Use \url without the optional [HREF]
%% argument when you want to print the url directly in the text. Otherwise,
%% use either \url or \anchor, with the HREF as the first argument and the
%% text to be printed in the second.

\acknowledgments

\clearpage

\clearpage

\end{document}